\documentclass[epj]{webofc}
\usepackage[varg]{txfonts}

\newif\iffigs\figstrue

\usepackage{epsfig,latexsym}
\usepackage{amsmath}
\usepackage{verbatim}
\usepackage{mathrsfs}
\usepackage{amssymb}

\wocname{EPJ Web of Conferences}
\woctitle{New Frontiers in Physics 2014}

\def\gz0{\gamma^{0}}

\def\g{\gamma}

\def\vf{\varphi}

\def\cA{{\cal A}}

\def\be{\begin{equation}}
\def\ee{\end{equation}}
\def\bea{\begin{eqnarray}}
\def\eea{\end{eqnarray}}
\def\ba{\begin{array}}
\def\ea{\end{array}}
\def\bec{\begin{center}}
\def\ec{\end{center}}
\def\ba{\begin{align}}
\def\ena{\end{align}}

\def\12{\frac{1}{2}}

\begin{document}
\pagestyle{empty}
\selectlanguage{english}

\begin{flushright}
CERN-PH-TH/2014-236\\
{\today}
\end{flushright}

\vspace{20pt}

\begin{center}


{\Large\sc String Theory clues for the low--$\ell$ CMB ?}\\


\vspace{38pt}
{\sc N.~Kitazawa${}^{\; a}$ and A.~Sagnotti${}^{\; b,*}$}\\[20pt]

{${}^a$\sl\small Department of Physics, Tokyo Metropolitan University \\ Hachioji, Tokyo 192-0397, JAPAN \\ }
\vspace{12pt}

{${}^b$\sl\small Department of Physics, CERN Theory Division\\
CH - 1211 Geneva 23, SWITZERLAND \\ }
\vspace{6pt}

\vspace{1.5cm}

{\sc\large Abstract}
\end{center}
  ``Brane Supersymmetry Breaking'' is a peculiar string--scale mechanism that can unpair Bose and Fermi excitations in orientifold models. It results from the simultaneous presence, in the vacuum, of collections of D-branes and orientifolds that are not mutually BPS, and is closely tied to the scale of string excitations. It also leaves behind, for a mixing of dilaton and internal breathing mode, \emph{an exponential potential that is just too steep for a scalar to emerge from the initial singularity while descending it}. As a result, in this class of models the scalar can generically bounce off the exponential wall, and this dynamics brings along, in the power spectrum, \emph{an infrared depression typically followed by a pre--inflationary peak}. We elaborate on a possible link between this type of bounce and the low--$\ell$ end of the CMB angular power spectrum. For the first 32 multipoles, one can reach a 50\% reduction in $\chi^{\,2}$ with respect to the standard $\Lambda$CDM setting.
\vskip 30pt
\noindent{\small \sl Based on talks presented by A.~S. at Planck 2014, Paris, May 26 -- 30 2014, PASCOS 2014, Warsaw, June 22 -- 27 2014, String Phenomenology 2014, Trieste, July 7 -- 11 2014, and at the 3rd international Conference on New Frontiers in Physics, Kolymbari, Crete, July 28 -- August 6 2014. \vskip 24pt
\begin{center} To appear in the Proceedings of ICNFP \ 2014 \end{center} }
\vfill
\line(1,0){250}\\
{\footnotesize {$^{*}$On leave of absence from Scuola Normale Superiore and INFN, Piazza dei Cavalieri 7, 56126 Pisa ITALY}}

\noindent

\setcounter{page}{1}

\pagebreak

\newpage

\wocname{EPJ Web of Conferences}
\woctitle{New Frontiers in Physics 2014}

\title{String Theory clues for the low--$\ell$ CMB ?}

\author{
N. Kitazawa\,\inst{1}\fnsep
\thanks{\email{kitazawa@phys.se.tmu.ac.jp}}
\and
A. Sagnotti\,\inst{2}\fnsep
\thanks{\email{sagnotti@sns.it}}
}

\institute{
Department of Physics, Tokyo Metropolitan University Hachioji, Tokyo 192-0397, JAPAN
\and
Department of Physics, CERN Theory Division,
CH - 1211 Geneva 23, SWITZERLAND\\
{\footnotesize (On leave from Scuola Normale Superiore and INFN, Piazza dei Cavalieri 7, 56126 Pisa ITALY)}
}

\abstract{%
   ``Brane Supersymmetry Breaking'' is a peculiar string--scale mechanism that can unpair Bose and Fermi excitations in orientifold models. It results from the simultaneous presence, in the vacuum, of collections of D-branes and orientifolds that are not mutually BPS, and is closely tied to the scale of string excitations. It also leaves behind, for a mixing of dilaton and internal breathing mode, \emph{an exponential potential that is just too steep for a scalar to emerge from the initial singularity while descending it}. As a result, in this class of models the scalar can generically bounce off the exponential wall, and this dynamics brings along, in the power spectrum, \emph{an infrared depression typically followed by a pre--inflationary peak}. We elaborate on a possible link between this type of bounce and the low--$\ell$ end of the CMB angular power spectrum. For the first 32 multipoles, one can reach a 50\% reduction in $\chi^{\,2}$ with respect to the standard $\Lambda$CDM setting.}

\maketitle
%


\section{\sc  Introduction}\label{sec:intro}

Despite the intricacies of their vacuum configurations and a number of vexing open questions, String Theory \cite{strings} and its low--energy Supergravity \cite{supergravity} provide a profound and concrete candidate framework for High--Energy Physics beyond the Standard Model. They add to it gravity, dark--matter candidates and clues on the origin of flavor, but in typical constructions the string scale $M_s \sim \frac{1}{\sqrt{\alpha'}}$, where $\alpha'$ is the string Regge slope, lies only a few orders of magnitude below the Planck scale, so that clear--cut signatures of String Theory appear largely out of reach for collider experiments or high--precision measurements. However, the large--scale Universe can host important relics of its very early epochs, when typical energies were far closer to the Planck scale.

The four--dimensional models of Particle Physics inspired by String Theory are relatively under control in the presence of several supercharges, but the breaking of Supersymmetry is fraught with technical and conceptual complications. Strictly speaking, if Supersymmetry plays a role in Nature, at most four supercharges can survive down to TeV scales \cite{LHC}. Among many options suggested by String Theory, ``brane supersymmetry breaking'' (BSB) \cite{bsb} is an appealing case study, since it combines relative simplicity and a certain degree of rigidity.
It is realized in some orientifold vacua \cite{orientifolds} where Ramond--Ramond (RR) charge neutrality requires the simultaneous presence of D-branes and orientifolds that are not mutually BPS. It results in a non--linear realization of Supersymmetry in the low--energy Supergravity \cite{10d_bsb_couplings} and, interestingly, does not introduce tree--level tachyon instabilities.

The breaking of Supersymmetry is a typical feature of inflationary models \cite{inflation} based on Supergravity \cite{sugra_inflation}, but BSB possesses a striking and distinctive signature. This is \emph{a steep exponential potential that the inflaton cannot descend when emerging from an initial singularity} but where it can generically experience a bounce. This remarkable feature, however, is strictly linked to the two--derivative terms of the low--energy Supergravity, and the key question is whether and how these terms might still dominate, near the initial singularity, despite the higher--derivative string corrections. It would seem unlikely, but we do not know a complete answer to this crucial question. However, the explicit analysis of second--order corrections in \cite{cd} showed that the Gauss--Bonnet combination, which intriguingly plays a role in String Theory \cite{z}, causes no obstruction. Let us therefore proceed to review this ``climbing phenomenon'' before turning to its potential application to the CMB.

To begin with, for the Sugimoto model \cite{bsb}, which provides the simplest manifestation of BSB, in the Einstein frame the two--derivative terms of the ten--dimensional effective action read
\be
  \mathcal{S} \,= \, \frac{1}{2\, k_N^2} \ \int \, d^{\,10} \,x \, \sqrt{- \, \mbox{det} \,g} \left[ \, R \, + \ \frac{1}{2}\ g^{\mu\nu} \, \partial_\mu \,\phi \ \partial_\nu \, \phi \, - \, T\, e^{\, \frac{3}{2}\ \phi} \ + \ \ldots \right] \ ,
  \label{scalar10}
\ee
where the ellipsis is a reminder that $\alpha^\prime$ and string corrections are left out, together with all fermionic couplings. Moreover, as shown in \cite{as13,fss}, after a reduction to $d$ dimensions and taking into account the breathing mode one is led to a similar expression,
\be
S_{d} \ = \ \frac{1}{2\kappa_{d}^2} \ \int d^{\,d} x \sqrt{-g}
\  \Big[  \, R \ + \ \frac{1}{2} \ (\partial \Phi_s)^2  \ + \ \frac{1}{2} \, (\partial \Phi_t)^2 \ -  \ T_{9}  \  e^{\, \sqrt{\frac{2(d-1)}{d-2}}\ \Phi_t} \ + \ \ldots \Big] \, , \label{st6}
\ee
where $\Phi_s$ and $\Phi_t$ are two different admixtures of dilaton and breathing mode. Remarkably, in all these cases the exponential potential is ``critical'' in the sense of \cite{dks}, as we now turn to explain.

This discussion rests on three ingredients. The first is the class of metrics
\be
ds^{\,2} \ = \ e^{\,2\,{\cal B}(t)} \, dt^2 \ - \ e^{\,\frac{2\,{\cal A}(t)}{d-1}}\ d{\bf x} \,\cdot \, d{\bf x} \ , \label{FLRW_gen}
\ee
which would be typical of inflationary models were it not for the ``gauge function'' ${\cal B}$. The second is a convenient gauge choice for ${\cal B}$,
\be
V \, e^{\,2\,{\cal B}} \ = \ V_0 \ , \label{gauge}
\ee
which is possible provided the potential never vanishes, as is the case for the exponentials in eqs.~\eqref{scalar10} and \eqref{st6}. The third ingredient is a convenient set of rescalings,
\be
\tau \ = \ t\, \sqrt{\frac{d-1}{d-2}} \ , \qquad \varphi \ = \ \Phi_t\, \sqrt{\frac{d-1}{2(d-2)}}  \qquad \left( \, \varphi \, =\,  \frac{3\, \phi}{4} \quad {\rm for \ \  d = 10\ } \right) \ , \label{varphiphit}
\ee
which have the virtue of casting the cosmological equations in the $d$--independent form
\bea
\qquad \qquad \qquad \dot{\cal A}^{\,2} \ - \ \dot{\varphi}^{\,2} & = & 1 \ , \nonumber \\
\qquad \qquad \qquad \ddot{\varphi} \ + \ \dot{\varphi} \sqrt{1 \ + \ \dot{\varphi}^{\,2}} \ + \ \frac{V_{\varphi}}{2\, V} \, \left( 1 \ + \ \dot{\varphi}^{\,2} \right) & = & 0 \ , \label{eqsgaugeB}
\eea
where ``dots'' stand for $\tau$ derivatives.

The solutions of this system if
\be\  \label{expotential}
V \ = \ V_0 \ e^{\,2\,\gamma\,\varphi}
\ee
have a long history \cite{lm,exp_sol,dm_vacuum,russo}, but it is fair to state that a key lesson was appreciated only recently \cite{dks}. One can derive their general form noting, as in \cite{dm_vacuum,russo}, that with the potential \eqref{expotential} the scalar equation is effectively of first order.
It is instructive, however, to begin from the Lucchin--Matarrese attractor (LM) \cite{lm}, which is characterized by a constant speed in this gauge, only exists for $\gamma<1$ and takes the suggestive form
\be
\dot{\varphi} \ = \ - \ \frac{\gamma}{\sqrt{1-\gamma^2}} \ . \label{lm_atrr}
\ee
The speed thus diverges as $\gamma$ approaches one from below, where the LM attractor disappears altogether. Now there is an intriguing fact: \emph{the Polyakov expansion of String Theory assigns to the potential emerging from ``brane supersymmetry breaking'' the value $\gamma=1$}, so that it sits precisely where the dynamical system of eqs.~\eqref{eqsgaugeB} exhibits a sharp change of behavior.

Referring for definiteness to the expanding case $\dot{\cal A}>0$, one can show that for $0< \gamma<1$ there are \emph{two} classes of solutions, which describe respectively a scalar that emerges from the initial singularity while \emph{climbing} or \emph{descending} the potential.
The \emph{climbing} solutions for the $\tau$--derivatives of $\vf$ and ${\cal A}$ are
\bea
\dot{\vf} &=& \frac{1}{2} \left[ \sqrt{\frac{1\,-\, \g}{1\,+\, \g}}\, \coth \left( \frac{\tau}{2}\ \sqrt{1\,-\, \g^{\,2}}\, \right) \ - \ \sqrt{\frac{1\,+\, \g}{1\,-\, \g}}\, \tanh
\left( \frac{\tau}{2}\ \sqrt{1\,-\, \g^{\,2}}\, \right)\right]\ , \nonumber \\
\dot{\cal A} &=& \frac{1}{2} \left[ \sqrt{\frac{1\,-\, \g}{1\,+\, \g}}\, \coth \left( \frac{\tau}{2}\ \sqrt{1\,-\, \g^{\,2}}\, \right) \ + \ \sqrt{\frac{1\,+\, \g}{1\,-\, \g}}\, \tanh
\left( \frac{\tau}{2}\ \sqrt{1\,-\, \g^{\,2}}\, \right)\right]
 \ , \label{speeds}
\eea
and the reader should appreciate that these expressions \emph{do not involve any initial--value constants} other than the Big--Bang time, here set for convenience at $\tau=0$. On the other hand, the corresponding fields read
\bea
\varphi  &=& \varphi_0 \ + \ \frac{1}{1+\gamma} \ \log \sinh  \left( \frac{\tau}{2}\ \sqrt{1\,-\, \g^{\,2}}\, \right) \ - \ \frac{1}{1-\gamma} \ \log \cosh  \left( \frac{\tau}{2}\ \sqrt{1\,-\, \g^{\,2}}\, \right)\ , \nonumber \\
{\cal A} &=& \ \frac{1}{1+\gamma} \ \log \sinh  \left( \frac{\tau}{2}\ \sqrt{1\,-\, \g^{\,2}}\, \right) \ + \ \frac{1}{1-\gamma} \ \log \cosh  \left( \frac{\tau}{2}\ \sqrt{1\,-\, \g^{\,2}}\, \right) \ .\label{fields_1exp}
\eea
They do involve an important \emph{integration constant}, $\varphi_0$, which places an upper bound on the values that $\varphi$ can attain during the cosmological evolution. Since in these models the inflaton hosts at least an admixture of the dilaton, the existence of an upper bound for a climbing scalar implies that \emph{the climbing behavior can be under control in string perturbation theory}. ${\cal A}$ would also involve an additive constant, but one can set it to zero up to a rescaling of the spatial coordinates. The solutions for a scalar that emerges from the initial singularity \emph{descending} the potential can be simply obtained from eqs.~\eqref{speeds} and \eqref{fields_1exp}
with the replacements $(\gamma,\varphi) \to (- \gamma,- \varphi)$,
which are a symmetry of the actions \eqref{scalar10} and \eqref{st6} with the potential \eqref{expotential}. All solutions converge for large $\tau$ to the LM attractor of eq.~\eqref{lm_atrr}.

To reiterate, for $\gamma \geq 1$ only the climbing solution exists, and remarkably the low--energy Supergravity \eqref{scalar10} for the ten--dimensional Sugimoto model in \cite{bsb} corresponds precisely to the critical value $\gamma=1$. Moreover, this property continues to hold for compactifications, insofar as the other scalar $\Phi_s$ is somehow stabilized by corrections that are implicitly present in eq.~\eqref{st6} and, as we have already cautioned, up to higher--derivative string terms \cite{cd}.

Keeping in mind these reservations, the exponential potential can convey a more general lesson. Namely, once the dynamics is dominated by a single scalar field, this cannot emerge from an initial singularity descending even more general potentials that are dominated asymptotically, say for $\varphi \to \infty$, by an exponential term with $\gamma \geq 1$. A fast bounce against a hard exponential potential would thus appear a generic fate, but by itself would not leave any tangible signs. However, if milder contributions were also present, the bounce could accompany the onset of an inflationary phase.

The simplest modification of eq.~\eqref{expotential} is provided by the two--exponential potentials
\be
{V}(\varphi) \ = \ {V}_0 \left( e^{\,2\,\varphi} \ + \ e^{\,2\,\gamma\, \varphi} \right) \ , \label{potwoexp}
\ee
with $\gamma < 1$, which were examined in detail in \cite{dks,dkps,ks}.
 These potentials have the virtue of combining the climbing phenomenon with the eventual approach of an LM attractor. Moreover, they never vanish and are thus naturally tailored for the gauge choice \eqref{gauge}, which goes a long way toward eliminating the stiffness of the dynamical system of eqs.~\eqref{eqsgaugeB}. General solutions are not available, but their main features can be anticipated from the single--exponential case. Moreover, qualitatively similar integrable potentials exist \cite{fss}, whose exact solutions clearly combine a fast climbing phase, a bounce and an eventual slow--roll phase. Finally, as we argued in \cite{as13,fss}, there are reasons to expect that milder exponentials emerge in String Theory. Indeed, under assumptions similar to the ones that led to eq.~\eqref{st6}, \emph{i.e.} provided $\Phi_s$ is somehow stabilized, a $p$--brane that couples to the dilation via $e^{\,-\, \alpha \,\phi}$ would yield a contribution like the second term in eq.~\eqref{potwoexp}, with
\be
\gamma \ = \ \frac{1}{12} \left( p \ + \ 9 \ - \ 6\, \alpha \right) \ .
\ee
All these values are quantized in units of $\frac{1}{12}$, a value that one could tentatively associate to an NS fivebrane wrapped along a small internal cycle, and which translates into a spectral index tantalizingly close to the PLANCK value $n_s \approx 0.96$ \cite{PLANCK}. To reiterate, the Polyakov expansion associates to BSB the hard exponential in eq.~\eqref{potwoexp}, up to the stabilization of $\Phi_s$, while its milder partner has potential links to other branes. Moreover, if $\gamma < \frac{1}{\sqrt{3}}$, even in the presence of other mild corrections $V^\prime(\varphi)$, a class of which we shall soon consider in a bottom--up approach, the bounce is followed by the onset of an inflationary phase.

This setting is clearly not fully realistic, since for one matter it does not address the exit problem, but \emph{our aim is more modestly to elaborate on the behavior of a climbing scalar near the turning point, under the (strong) assumption that its effects be accessible to us via the low--$\ell$ CMB}. Therefore, in the following we shall try to explore the potential signatures of this type of dynamics, under the spell of recent WMAP9 and PLANCK measurements of CMB power spectra of scalar perturbations \cite{WMAP9,PLANCK,PLANCK2}. Their low--$\ell$ ends display in fact some mild discrepancies with respect to the general attractor behavior $P(k) \sim k^{n_s-1}$ predicted long ago in \cite{cm}, of which PLANCK gave ample evidence. Let us stress that our considerations, if of any value, might prove of interest in two types of scenarios:
\begin{enumerate}
\item in the presence of a relatively small number of $e$--folds, not exceeding $N \sim 60$, if
 scalar perturbations generated during the very early stages of inflation happened to be entering the horizon at the present epoch. This would be most exciting, since we would be facing shadows of the initial singularity, even if inflation would perhaps loose some of its aesthetic appeal;
\item alternatively, and more conservatively, if the bounce occurred during a later transition between a fast--roll epoch and a subsequent slow--roll one.
\end{enumerate}

We can now turn to the indications of the climbing phenomenon for scalar CMB perturbations
once $\varphi$ is identified with an inflaton field. The following section rests on \cite{dkps}, where the infrared lack of power present in these types of scenarios was identified, but even more on the recent results in \cite{ks}, where the pre--inflationary peak resulting from the bounce was clearly exhibited. We shall also present some new results on improved power spectra and corresponding CMB fits, some of which were described in preliminary form by one of us during the Summer of 2014 \cite{as_14}.

\section{\sc  Pre--inflationary peaks in power spectra}\label{sec:powerspectrum}

In this section we would like to review briefly the content of \cite{dkps,ks} as a prelude to the tentative comparisons of the next section, while also including some new results. To begin with, the power spectrum of scalar perturbation $P_\zeta(k)$ is defined as
\be
\langle \zeta({\bf x},\tau_F)\,  \zeta({\bf x},\tau_F) \rangle \ = \ \int_0^\infty \frac{dk}{k} \ P_\zeta(k) \ , \label{power_zeta}
\ee
where $\zeta({\bf x},\tau)$ is the gauge invariant scalar perturbation field that is conserved outside the horizon
and $\tau_F$ is the ``parametric time'' at the end of inflation, after a large number of $e$--folds. Rather than $\zeta({\bf x},\tau)$, it is often convenient to use the Mukhanov--Sasaki (MS) variable
\be
v({\bf x},\tau) \ = z(\tau) \, \zeta({\bf x},\tau) \ , \qquad
{z}(\tau) \ = \ \frac{1}{k_N}\ \sqrt{6}\ e^{\,\frac{{\cal A}(\tau)}{3}} \ \frac{d\varphi(\tau)}{d{\cal A}(\tau)} \ . \label{zeta}
\ee
This has the virtue of not being singular at the bounce, and moreover its Fourier coefficients evolve, in conformal time $\eta$, according to the Schr\"odinger--like equation
\be
\left(\frac{d^2}{d \eta^2} \ + \ k^2 \ - \ W_s(\eta) \right) v_k(\eta) \ = \ 0\ ,
\label{MS-eq}
\ee
on which one can gain some insight via standard tools of Quantum Mechanics. Here
\be
d \eta \ = \ e^{\,-\,\frac{{\cal A}}{3}} \, \sqrt{\frac{V_0}{V}} \, d \tau \ , \qquad W_s \ = \ \frac{z^{\,\prime\prime}(\eta)}{z(\eta)} \ , \label{MS_potential}
\ee
with $\eta<0$, and $\tau \rightarrow \infty$ corresponds to $\eta \rightarrow 0^-$. The ``MS potential'' $W_s$ reflects the behavior of the background fields $\varphi$ and $\cA$, ``primes'' denote derivatives with respect to $\eta$ and $z$ is defined in eq.~\eqref{zeta}.

The solution of eq.~\eqref{MS-eq} should satisfy the Wronskian constraint
\be
v_k \, \frac{\partial}{\partial \, \eta} \ v_k^\star \ - \ v_k^\star \, \frac{\partial}{\partial \, \eta} \ v_k \ = \ i \ ,
\ee
and we follow standard practice in adopting the Bunch--Davies condition
\be
v_k(\eta) \ \ \ \thicksim\!\!\!\!\!\!\!\!_{{}_{{k \to \infty}}} \ \ \frac{1}{\sqrt{2k}}\, e^{\,-\,i\,k\,\eta}\ \ .
\ee
This natural assumption plays an important role, and would perhaps deserve further thoughts. The swamping of infrared effects that results from abandoning it was discussed, in this context, in \cite{dkps}. With this proviso, the determination of the power spectra of scalar perturbations via
\be
P_\zeta(k) \ = \ \frac{k^3}{2\,\pi^2} \ \left| \frac{v_k(-\epsilon)}{z(-\epsilon) }\right|^2  \label{power_spec}
\ee
results from an (initial--value) problem that bears some formal analogies to one--dimensional potential scattering in non--relativistic Quantum Mechanics.

The quantities involved are to be computed at the end of inflation, for small positive values of $\epsilon$, when the ratio becomes independent of $\epsilon$ and reduces to a well--defined function of $k$. Referring to the two--exponential model of eq.~\eqref{potwoexp}, from the limiting behavior of the background fields near the Big-Bang singularity and at late times one can conclude that
\be
W_s \ \ \ \thicksim\!\!\!\!\!\!\!\!\!\!_{{}_{{\eta \to - \eta_0}}} \ - \ \frac{1}{4}\ \frac{1}{(\eta+\eta_0)^2} \ , \qquad
W_s \ \ \ \thicksim\!\!\!\!\!\!\!\!\!\!_{{}_{{\eta \to - 0^-}}} \  \ \frac{\nu^2 - \frac{1}{4}}{\eta^2} \qquad \left( \nu \ = \ \frac{3}{2} \ \frac{1\,-\,\gamma^2}{1\,-\,3\,\gamma^2} \right) \  , \label{attractor_MS}
\ee
The initial singularity thus translates in a singular attractive behavior for the MS potential, while the final inflationary epoch builds up a ``centrifugal'' barrier. The mere presence of the former brings along a $k^3$--like \emph{infrared depression} of the power spectrum \eqref{power_spec}: this can be argued, for instance, by the simple WKB--like arguments presented in \cite{dkps}.
On the other hand, the barrier gives rise to a large \emph{amplification} of initial signals, which lies at the heart of a slight tilt of the power spectrum of the form
\be
P_\zeta(k) \ \sim \ k^{3-2\,\nu} \ \equiv \ k^{n_s-1}  \label{cibmukh}
\ee
predicted in \cite{cm} for the standard slow--roll inflation. The scalar evolution leaves its imprinting in the intermediate region, with power spectra that are somewhat similar but whose detailed features depend strongly on the value of $\varphi_0$.

\begin{figure}[h]
\begin{center}$
\begin{array}{ccc}
\epsfig{file=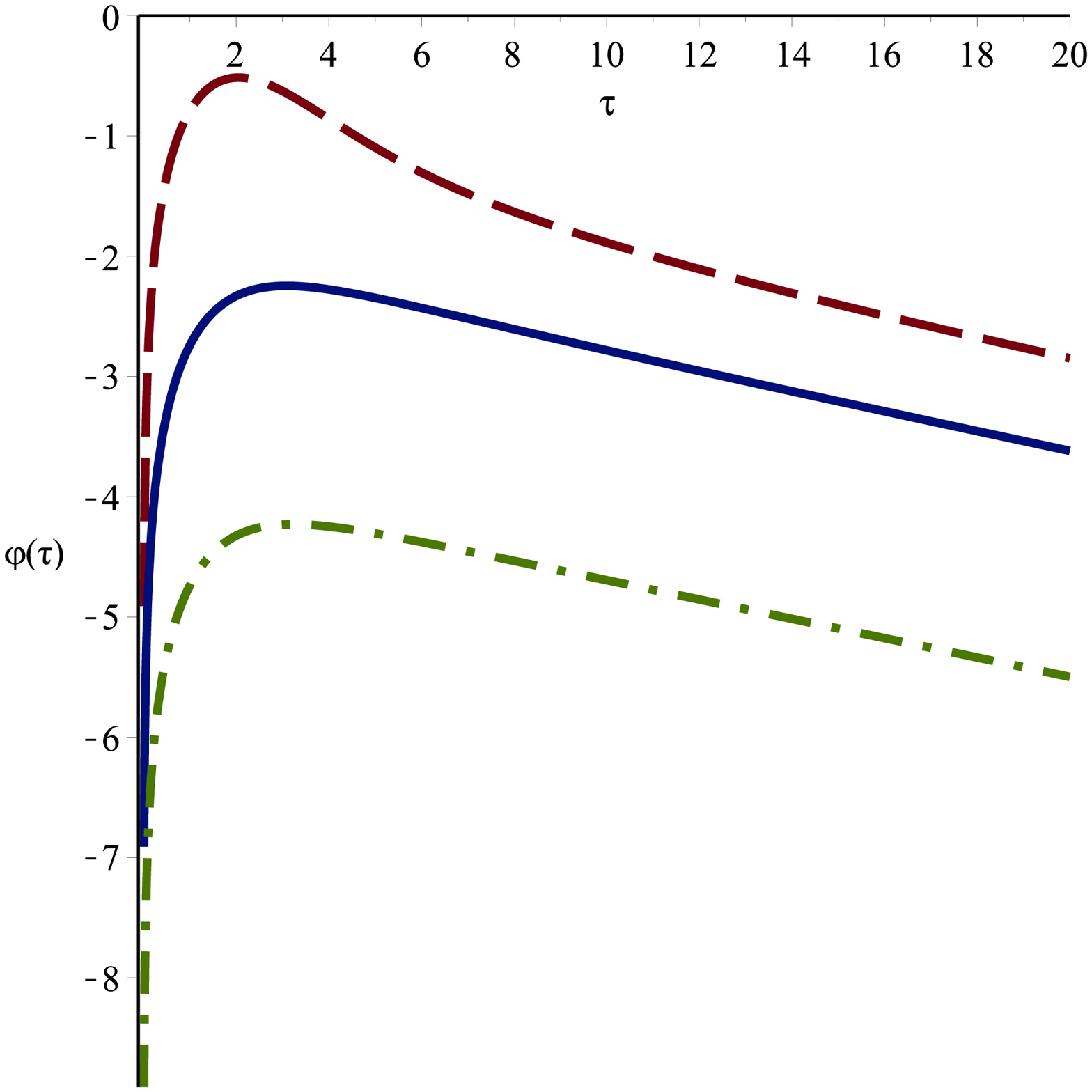, height=1.1in, width=1.1in} & \qquad\qquad
\epsfig{file=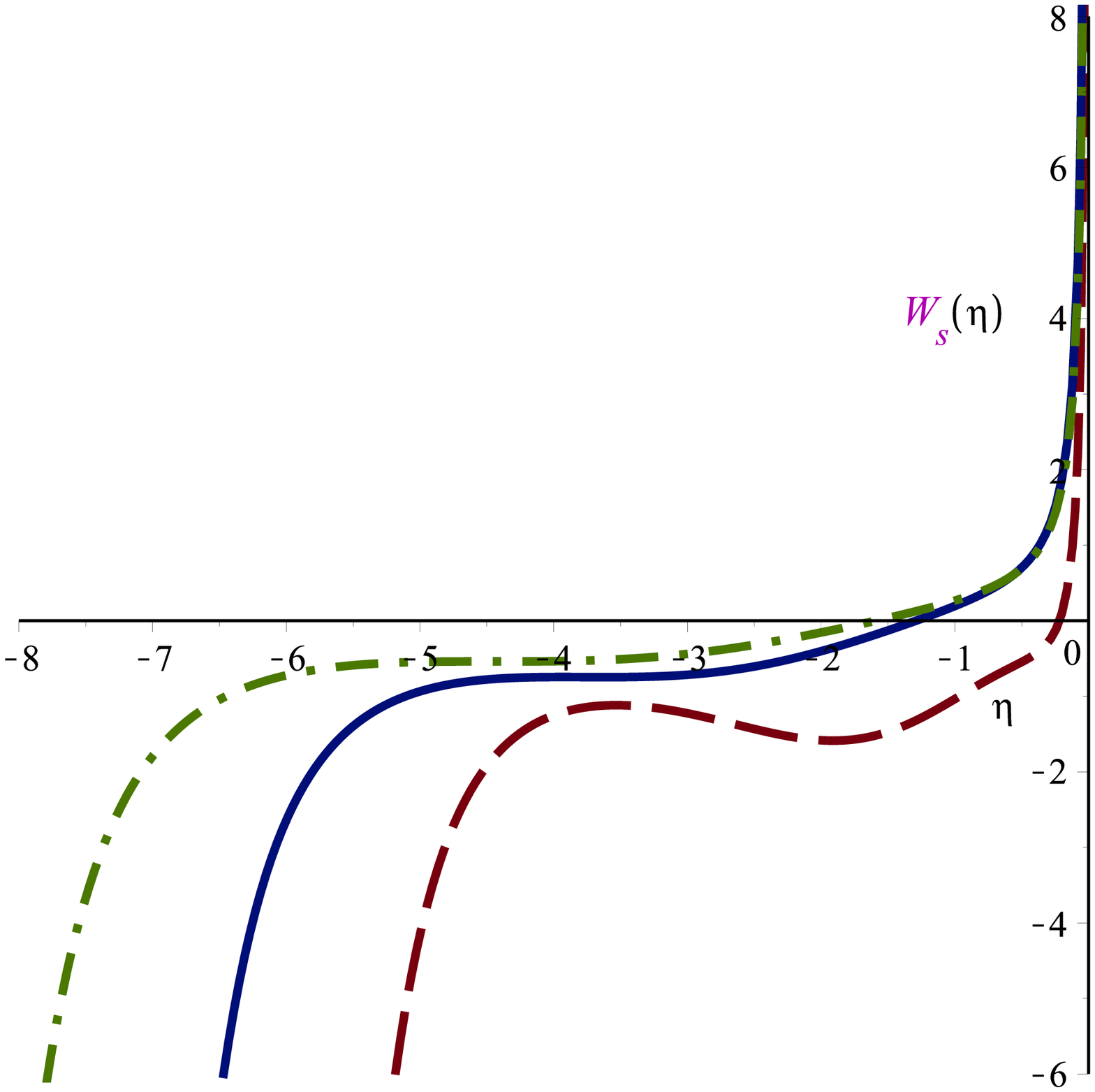, height=1.1in, width=1.1in} & \qquad\qquad
\epsfig{file=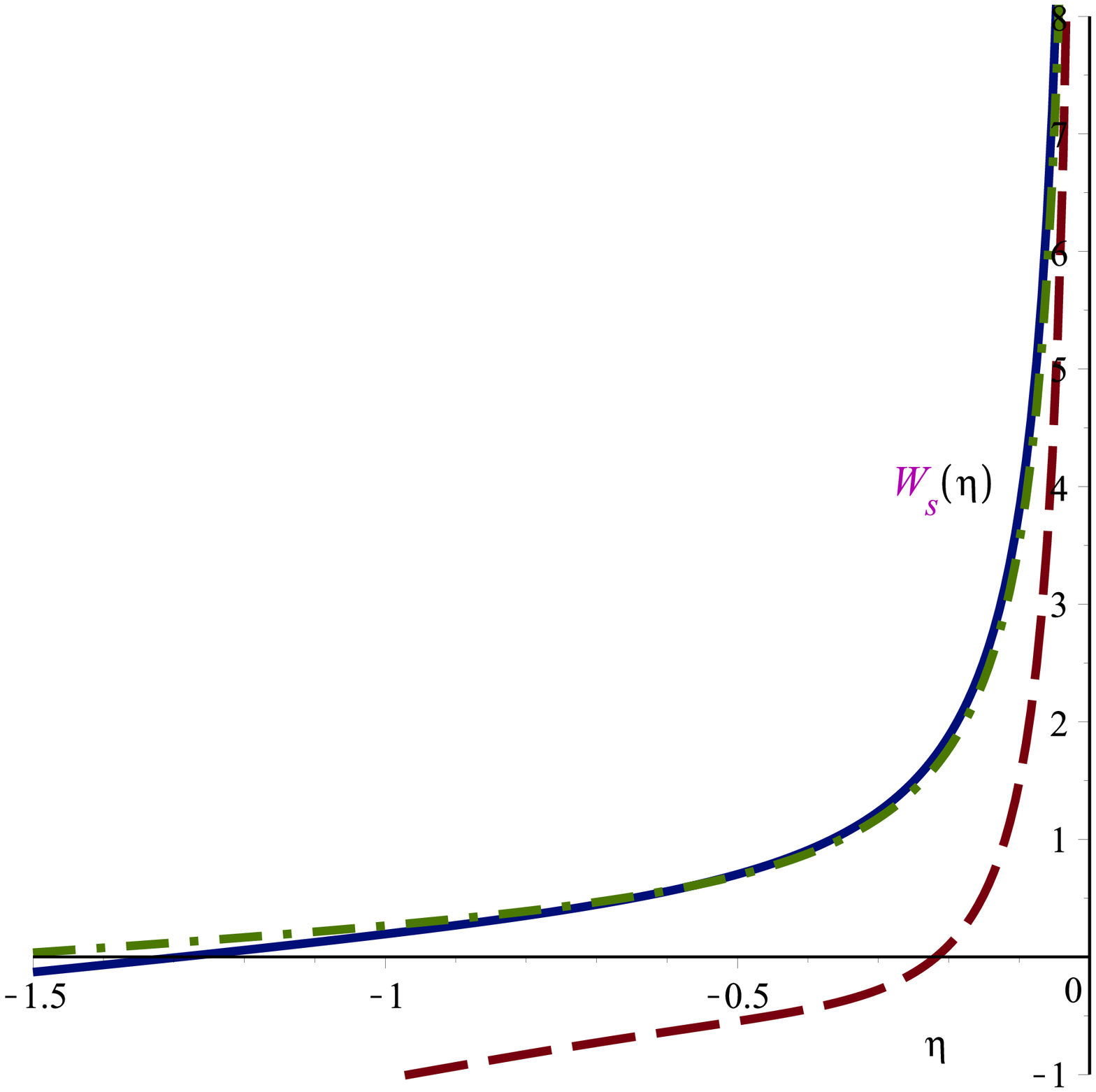, height=1.1in, width=1.1in}
\end{array}$
\end{center}
\caption{\small
$\tau$--evolution of the scalar field near the turning point (left),
 the corresponding MS potentials $W_s$ in conformal time $\eta$ (center)
 and an enlarged view of the region where the $W_s$ cross the horizontal axis (right).
 The lines for $\varphi_0=0,-2,-4$ are dashed, continuous and dashed--dotted.
}
\label{fig:phi_Ws}
\end{figure}

The results of some indicative evolutions in the two--exponential potential for $\varphi_0=0,-2,-4$ are shown in fig.~\ref{fig:phi_Ws}.
Notice that for larger values of $\varphi_0$, as the scalar climbs the potential further more complicated features emerge in $W_s$, but they are largely confined to its negative portion. The power spectra of scalar perturbations, however, are strongly affected by the way the system approaches the slow--roll epoch.

\begin{figure}[h]
\begin{center}$
\begin{array}{ccc}
\epsfig{file=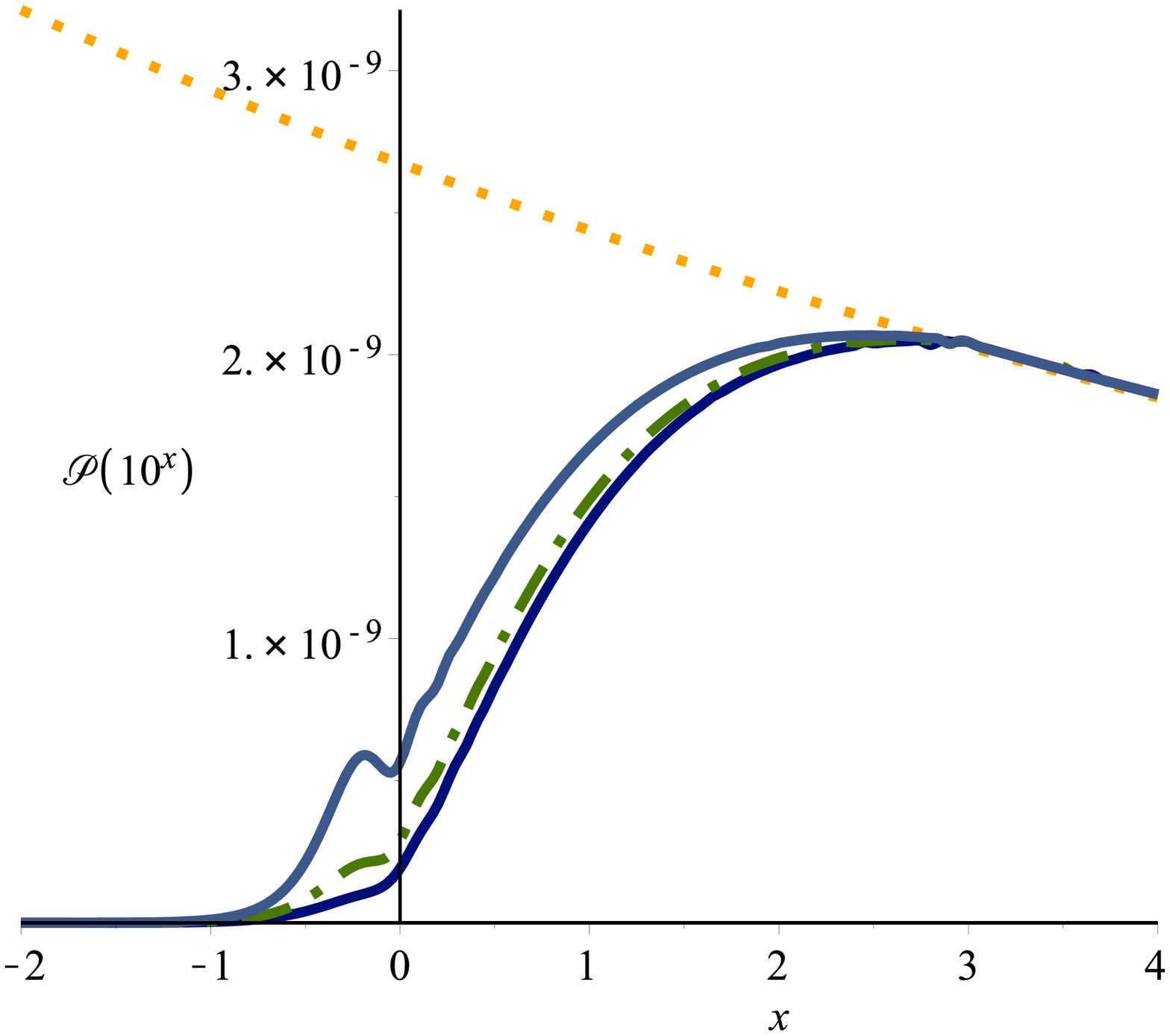, height=1.1in, width=1.1in} & \qquad\qquad
\epsfig{file=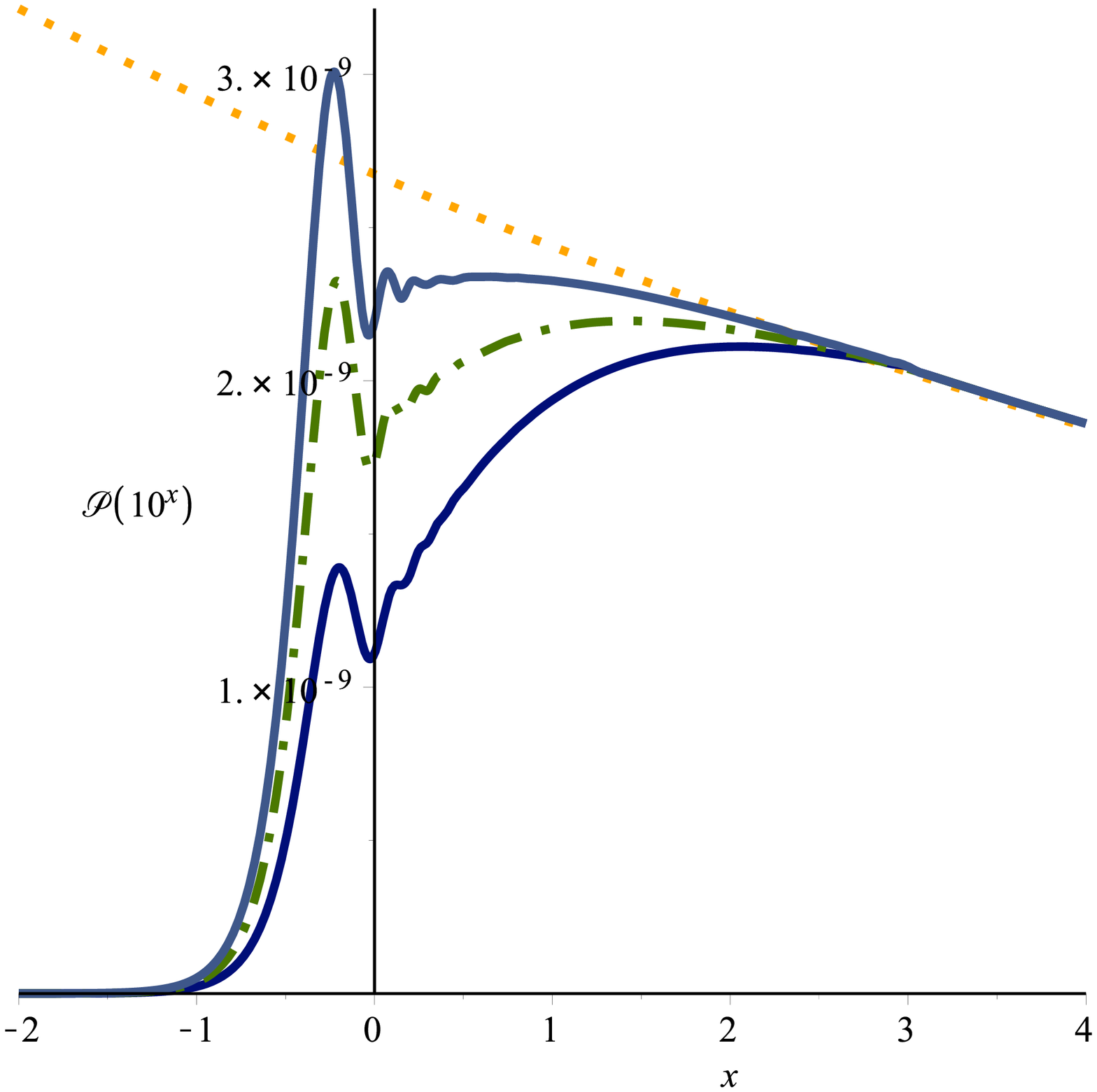, height=1.1in, width=1.1in} & \qquad\qquad
\epsfig{file=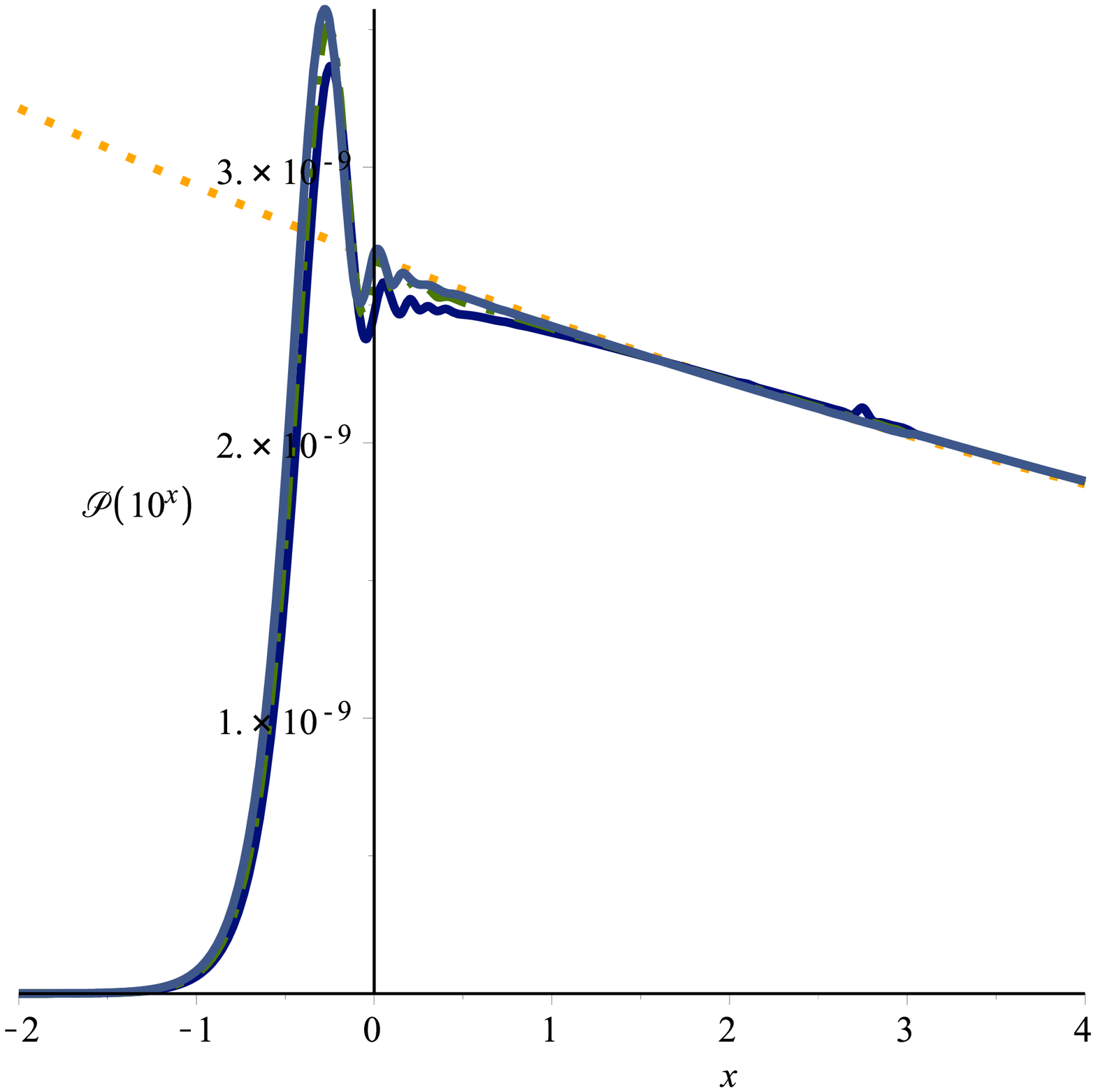, height=1.1in, width=1.1in}
\end{array}$
\end{center}
\caption{\small
Power spectra of scalar perturbations for the double--exponential potential of eq.~\eqref{potwoexp}
 (in all cases the dotted line is the attractor curve).
For $\varphi_0=0$ a featureless power spectrum approaches the attractor curve after four decades in $k$ (left),
 while for $\varphi_0=-0.5,-1$ a small pre--inflationary peak starts to build up (left).
For $\varphi_0=-1.5,-2,-2.5$ the pre--inflationary peak becomes more and more pronounced,
 but remains well separated from the attractor curve (center).
For $\varphi_0=-3,-3.5,-4$ the power spectrum rises steeply,
 and a narrow peak overshoots the attractor curve that is readily reached after a few oscillations (right).
On the horizontal axis we display $x$, where $k=10^x$,
 and these power spectra are normalized to meet at the end of the explored range.
}
\label{fig:double}
\end{figure}

Three families of power spectra well captured by nine significant choices of $\varphi_0$ are collected in Fig.~\ref{fig:double}.
For larger values of $\varphi_0$ there are no features, aside from the slow convergence to the attractor behavior discussed in detail in \cite{dkps}. This can be explained noting that the system is in fast-roll during the climbing phase and bounces relatively fast against the exponential wall, which leaves no tangible signs.
On the other hand, for small values of $\varphi_0$
the power spectra converge rapidly to the asymptotic behavior well described in \cite{destri}.

Mechanisms for the infrared depletion of power spectra were recently examined by several authors \cite{quadrupole_red}. Our picture rests on the distinctive signature of the climbing scalar, whose bounce against the exponential wall results in an infrared depression, followed by \emph{a pre--inflationary peak} and by a subsequent growth reaching up eventually the power--like attractor behavior.

\begin{figure}[h]
\begin{center}$
\begin{array}{cccc}
\epsfig{file=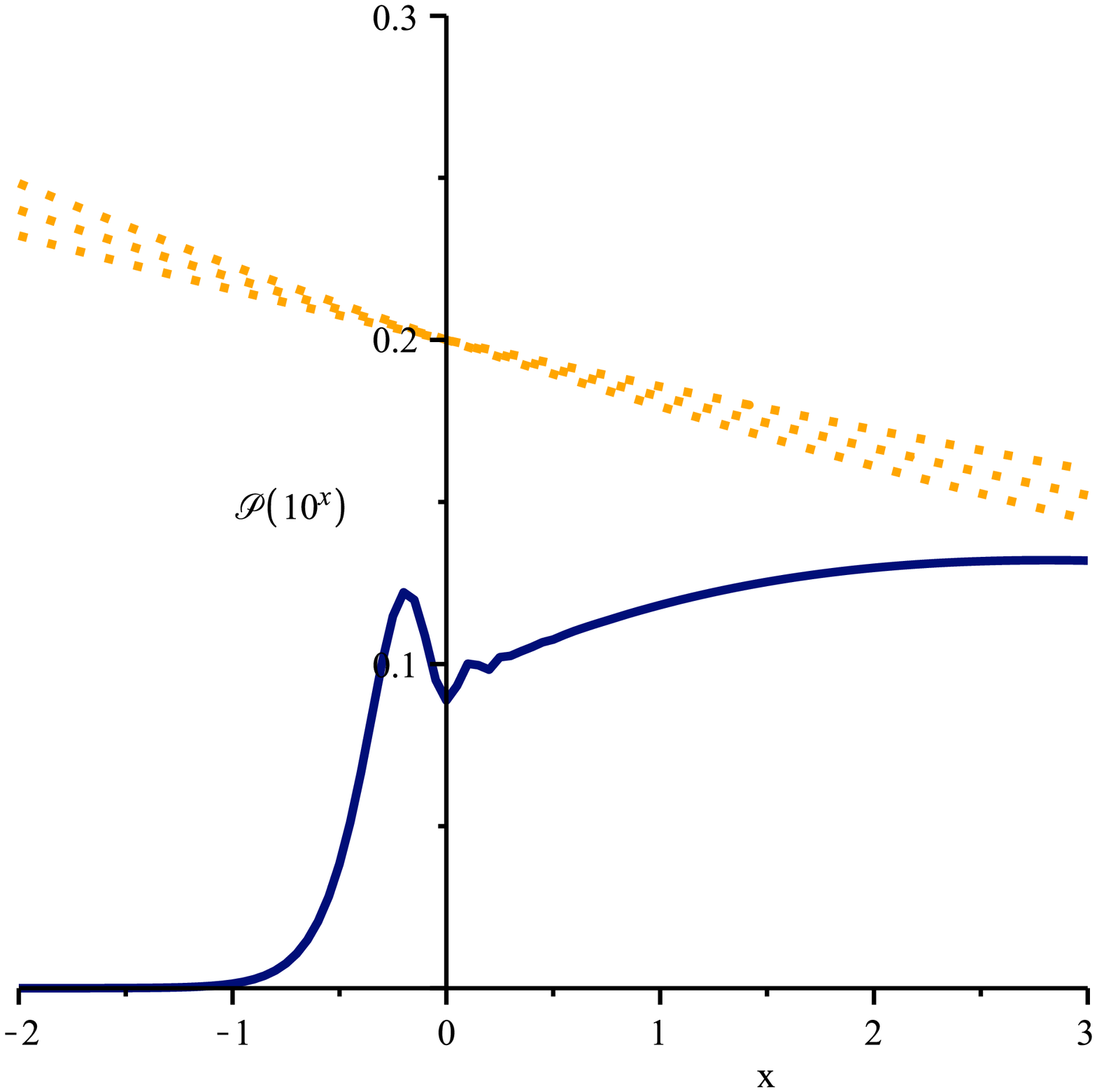, height=1in, width=1in}& \qquad
\epsfig{file=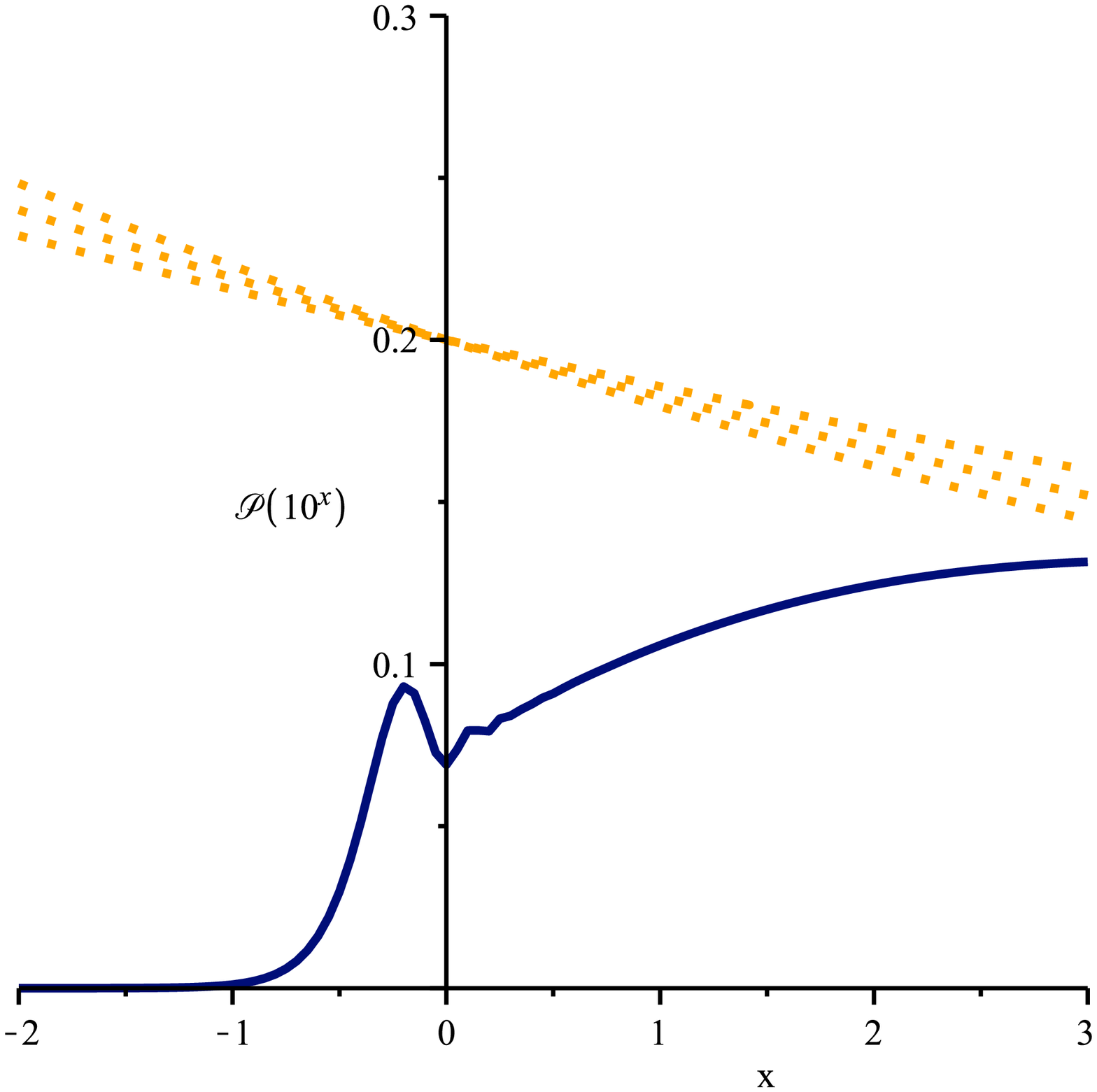, height=1in, width=1in}& \qquad
\epsfig{file=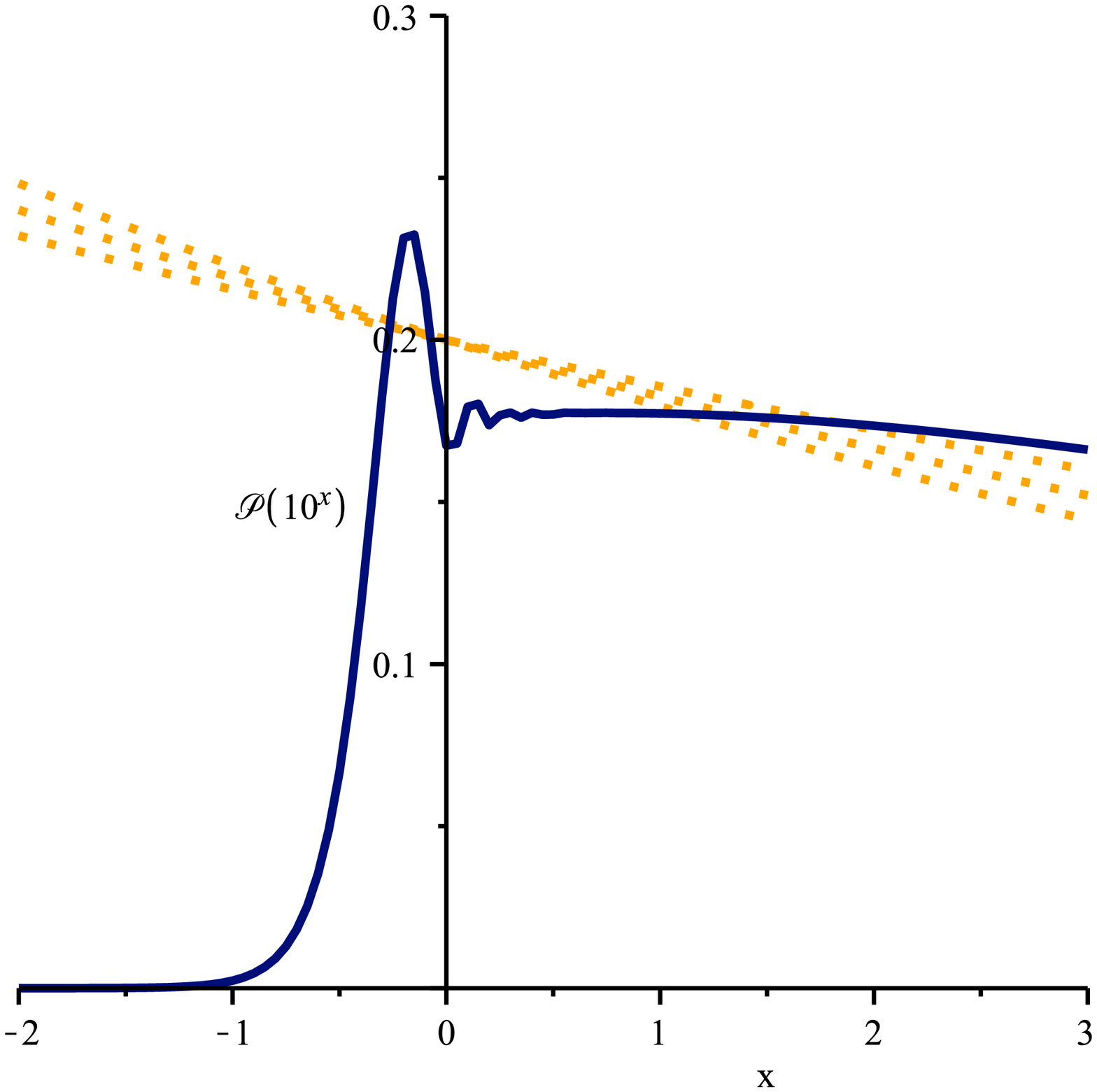, height=1in, width=1in}& \qquad
\epsfig{file=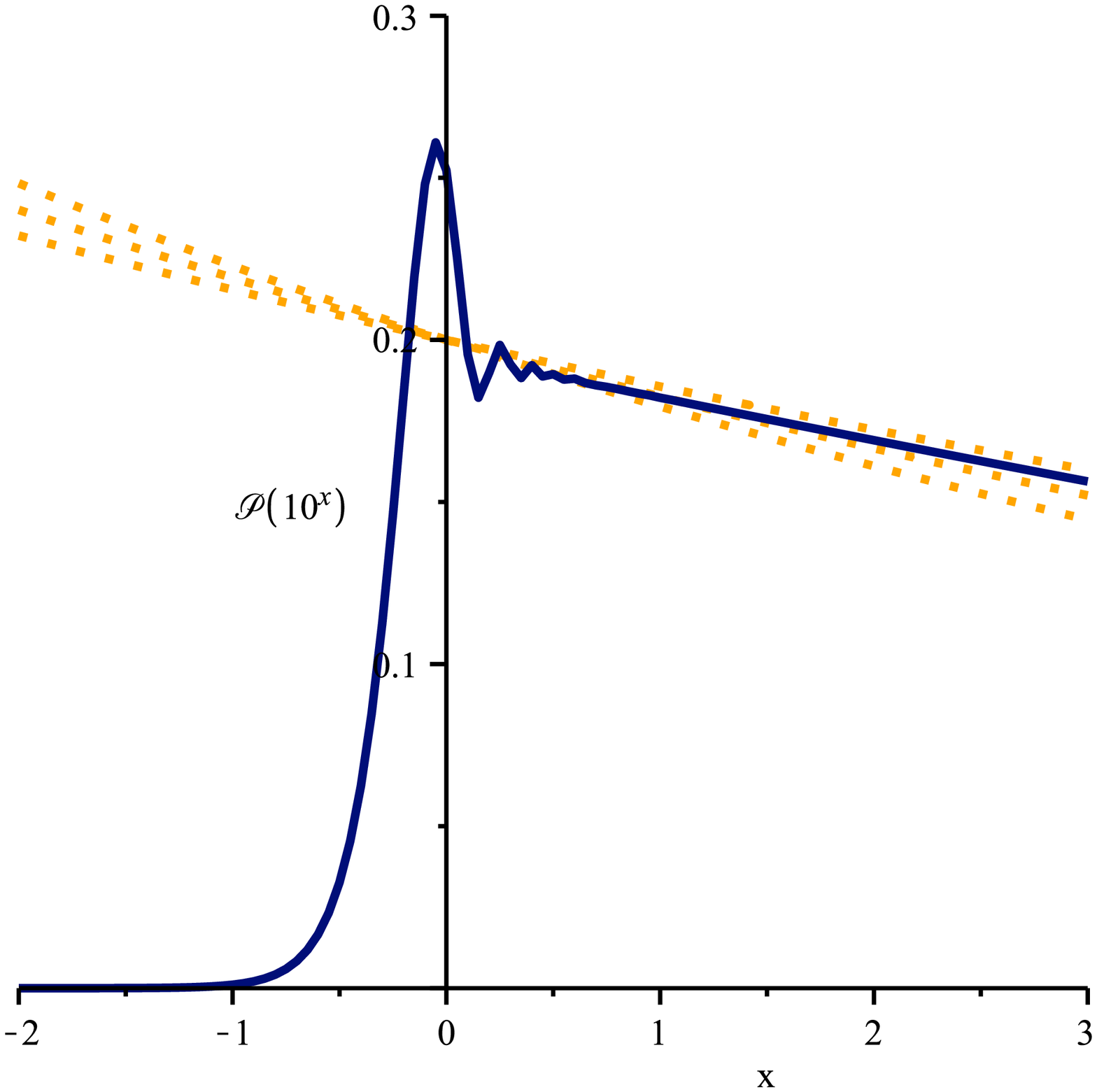, height=1in, width=1in}
\end{array}$
\end{center}
\caption{\small Power spectra of scalar perturbations computed, in cosmic time, for the Starobinsky--like potentials of eq.~\eqref{starobinsky}, shown here with an arbitrary normalization but with the parameters adjusted in order to guarantee about 60 $e$--folds of inflation and a fair portion of them with $n_s \simeq 0.96$. The slopes of the dotted lines reflect a range of values for $n_s$ that is close to current observations.}
\label{fig:starobinsky_power}
\end{figure}

The pre--inflationary peak emerges as the system approaches the exponential wall more slowly at the end of the climbing phase, so that the scalar perturbations that accompany the reversal of its motion can grow to some extent, while after the bounce the system has not quite reached the attractor. The pre--inflationary peak thus reflects the degree of slow--roll attained by the scalar field during this epoch, and the simple two--exponential model has the virtue of eliciting this general feature of the hard wall in a relatively simple context. However, similar spectra continue to arise in more realistic models, and for instance in Starobinsky--like potentials with a hard exponential, for which \cite{ks}
\be
{V}_S(\varphi) \ = \ {V}_0\, \left(1- e^{\,-\,\gamma\,(\varphi+\Delta)}\right)^2 \ + \ V_1 \, e^{\,2\,\varphi} \ , \label{starobinsky}
\ee
as shown in fig.~\ref{fig:starobinsky_power}. The reader should note that the potential is typically very flat, in this class of models, close to the exponential wall.

The examples that we have displayed for the two--exponential model were obtained with $\gamma~\sim~0.08~\sim~\frac{1}{12}$, which results in a spectral index of about 0.96 in the attractor region. However, the example of the Starobinsky potential is meant to stress, in compliance with the indications of PLANCK \cite{PLANCK}, the relevance of \emph{concave potentials} that turn gradually to the slope associated with the observed spectral index. Indeed, a slight preference this type of behavior also surfaced from our $\chi^2$--analysis in \cite{ks}, where the results improved slightly for lower values of $\gamma$.

\begin{figure}[h]
\begin{center}$
\begin{array}{ccc}
\epsfig{file=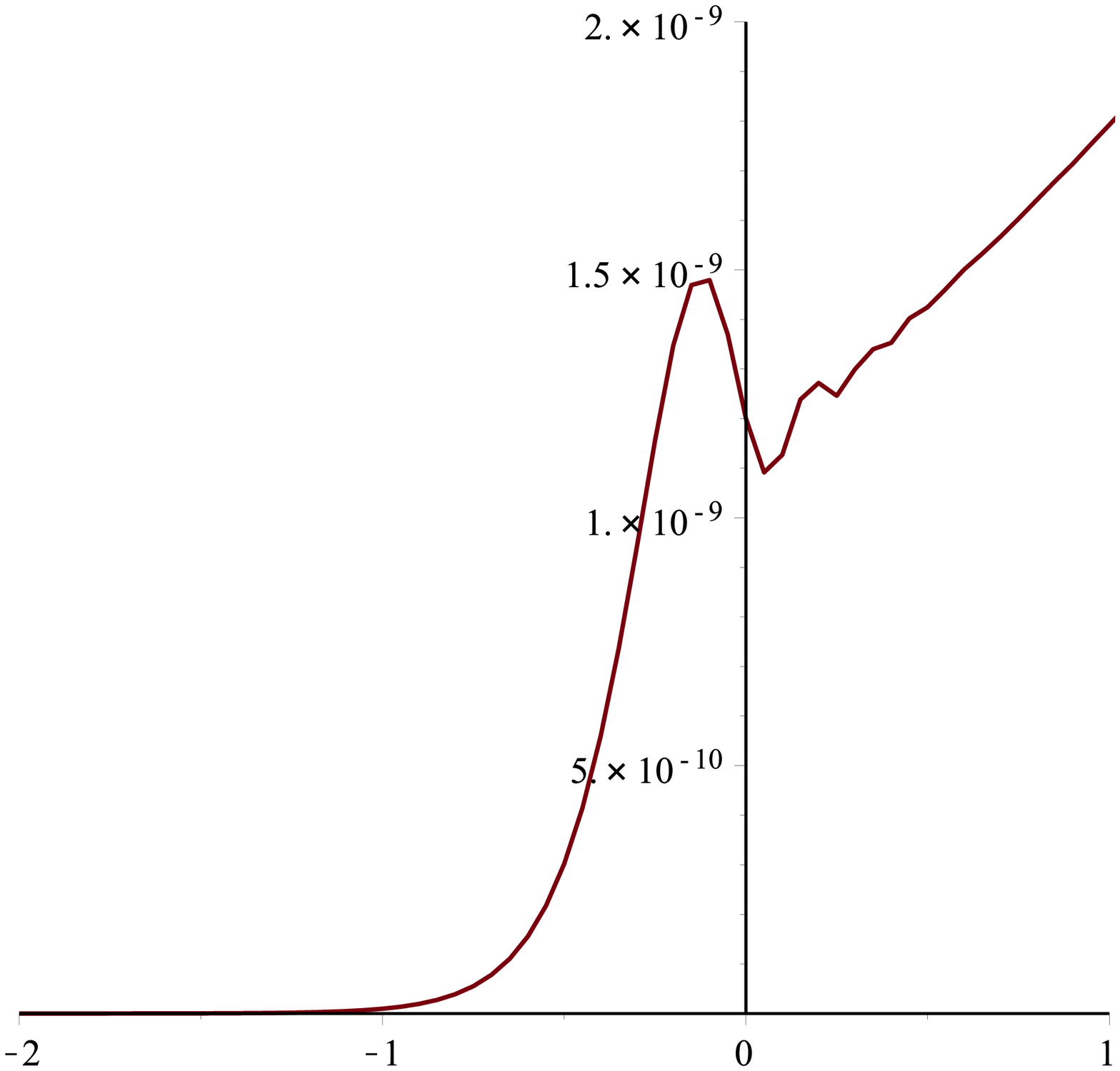, height=1.1in, width=1.1in} & \qquad\qquad
\epsfig{file=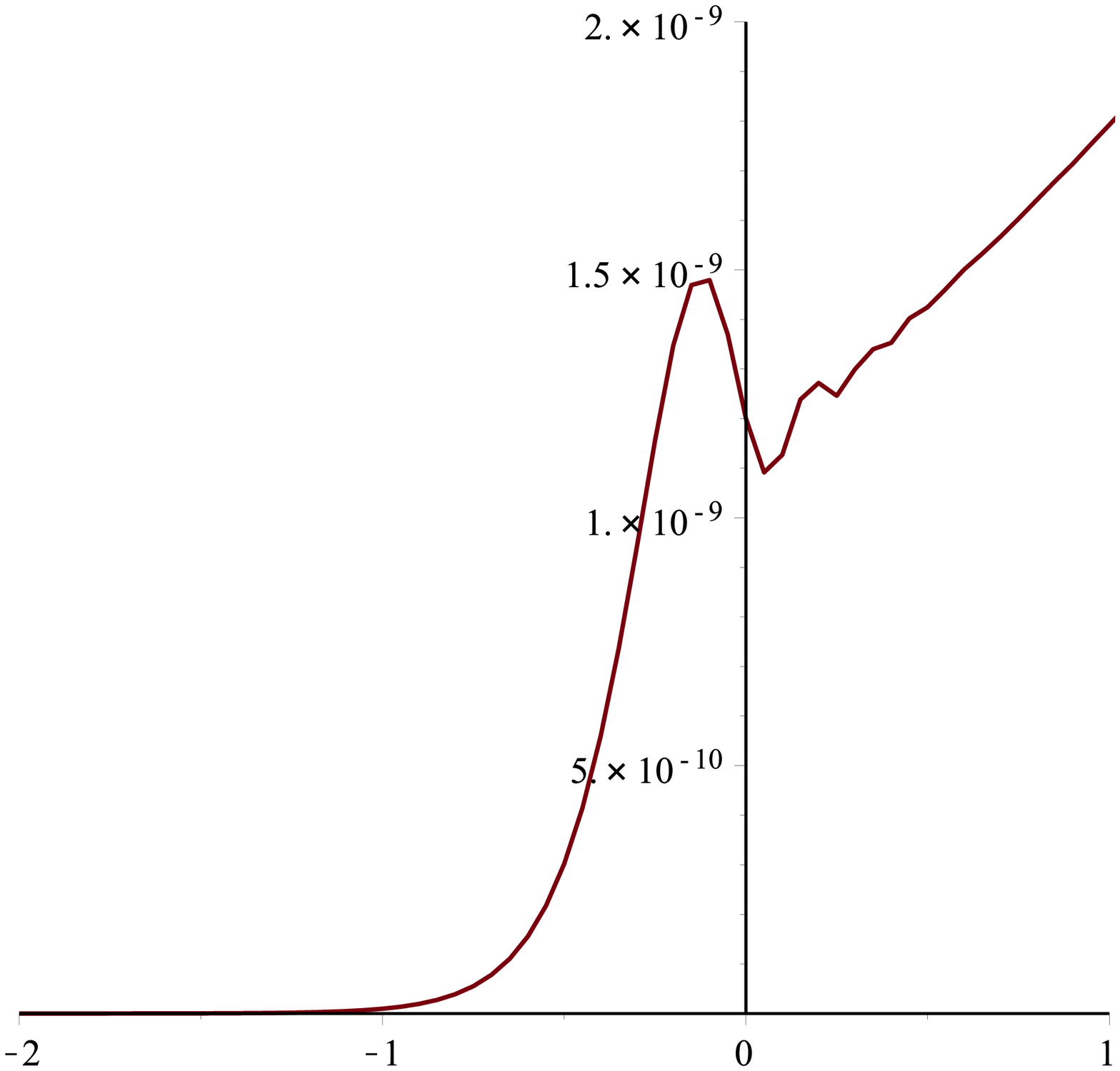, height=1.1in, width=1.1in} & \qquad\qquad
\epsfig{file=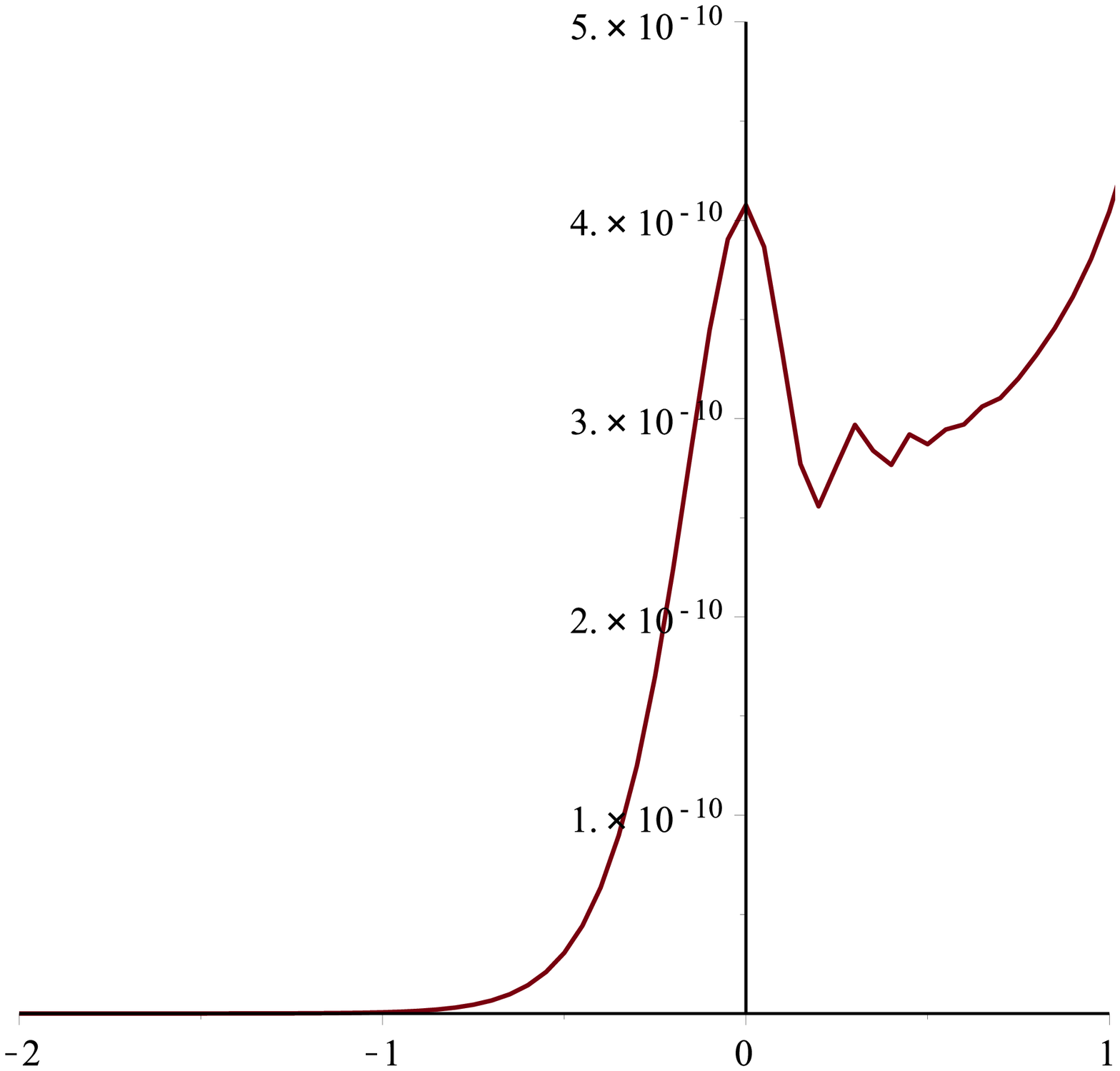, height=1.1in, width=1.1in}
\end{array}$
\end{center}
\caption{\small The effect of a slightly negative $\gamma$ in the two--exponential potential. We focus on a smaller portion of the $k$ that is still realistic with these models, and yet suffices to get fair $\chi^2$ estimates. The peaks are slightly enhanced and, as in the left portion of fig.~\ref{fig:double}, they are followed by steeper profiles that are initially convex.
}
\label{fig:negative_gamma}
\end{figure}

In closing this section, we would like to call to the reader's attention an interesting mechanism to enhance the pre--inflationary peaks. The idea is simple: slowing down slightly the scalar after the bounce favors the growth of perturbations arising from that region. One can foresee the effect working with slightly negative values of $\gamma$ in the potentials of eq.~\eqref{potwoexp}. While this option is admittedly somewhat artificial, it clearly results in \emph{more prominent pre--inflationary peaks followed by rapidly growing power spectra that start out slightly convex after the peaks}, as shown in fig.~\ref{fig:negative_gamma}. This is to be contrasted with the \emph{milder concave profiles} displayed in fig.~\ref{fig:double}, which characterize the $k$--dependence of the power spectrum beyond the pre--inflationary peak in standard two--exponential models. As we shall see, the low--$\ell$ CMB angular power spectrum appears to favor this type of power spectra of scalar perturbations.

\begin{figure}[h]
\begin{center}$
\begin{array}{cc}
\epsfig{file=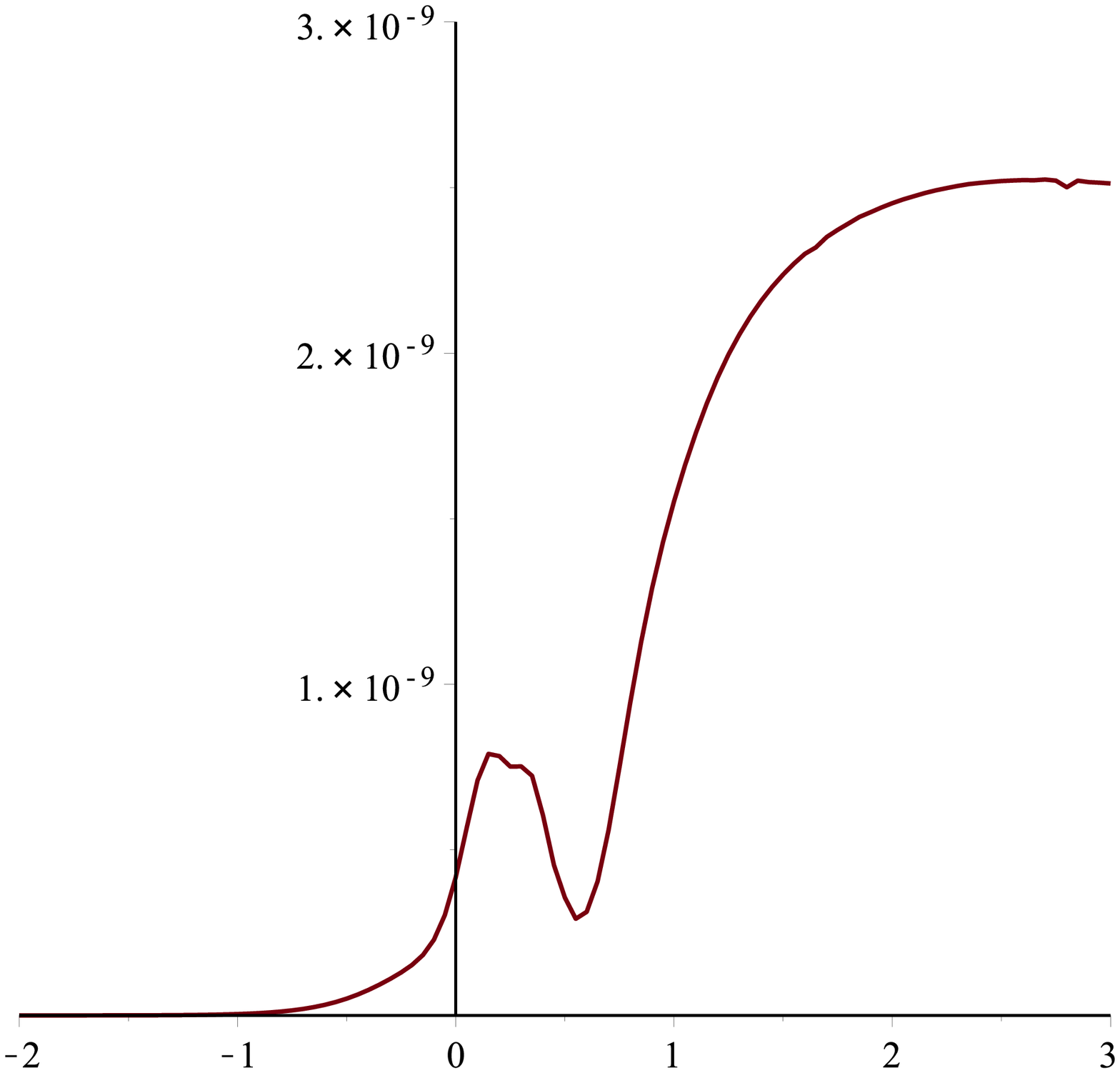, height=1.1in, width=1.1in} & \qquad\qquad
\epsfig{file=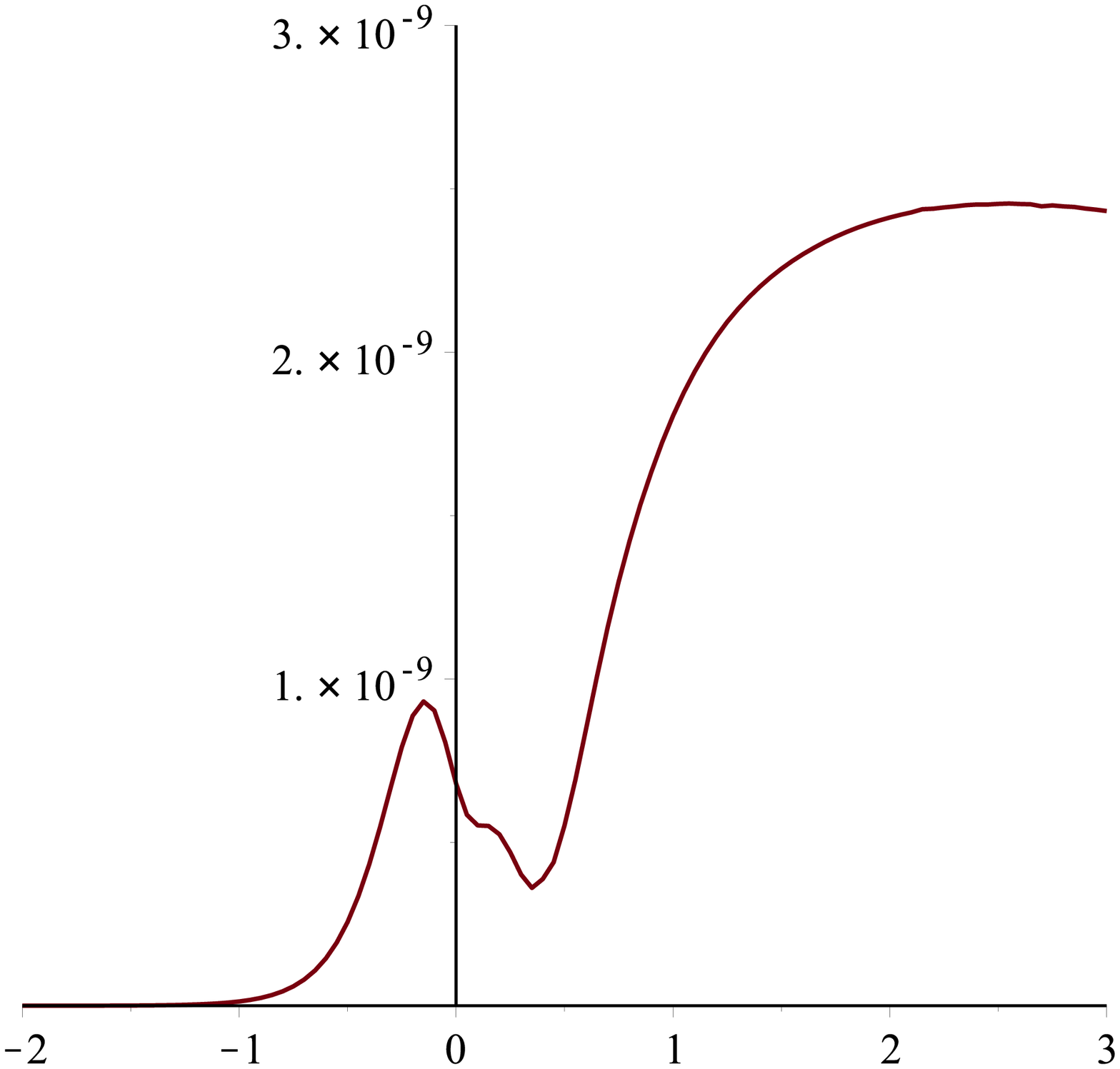, height=1.1in, width=1.1in}
\end{array}$
\end{center}
\caption{\small The effects of a small gaussian bump close to the exponential barrier of BSB. Notice the analogies with fig.~\ref{fig:negative_gamma}.
}
\label{fig:gaussian}
\end{figure}

More realistic scenarios where the scalar is slowed down right after the bounce obtain adding to the two--exponential potentials of eq.~\eqref{potwoexp} a small gaussian bump of the form
\be
{V}^\prime(\varphi) \ = \ {V}_0 \ a_1 \, e^{\,-\, a_2 \, \left(\varphi+a_3\right)^2} \ . \label{potwoexp_gauss}
\ee
Two examples of power spectra that result from this modification are collected in fig.~\ref{fig:gaussian}.
%
\section{\sc  $\chi^{\,2}$ analysis of the low--$\ell$ CMB angular power spectrum}\label{sec:observables}

\begin{figure}[h]
\begin{center}$
\begin{array}{ccc}
\epsfig{file=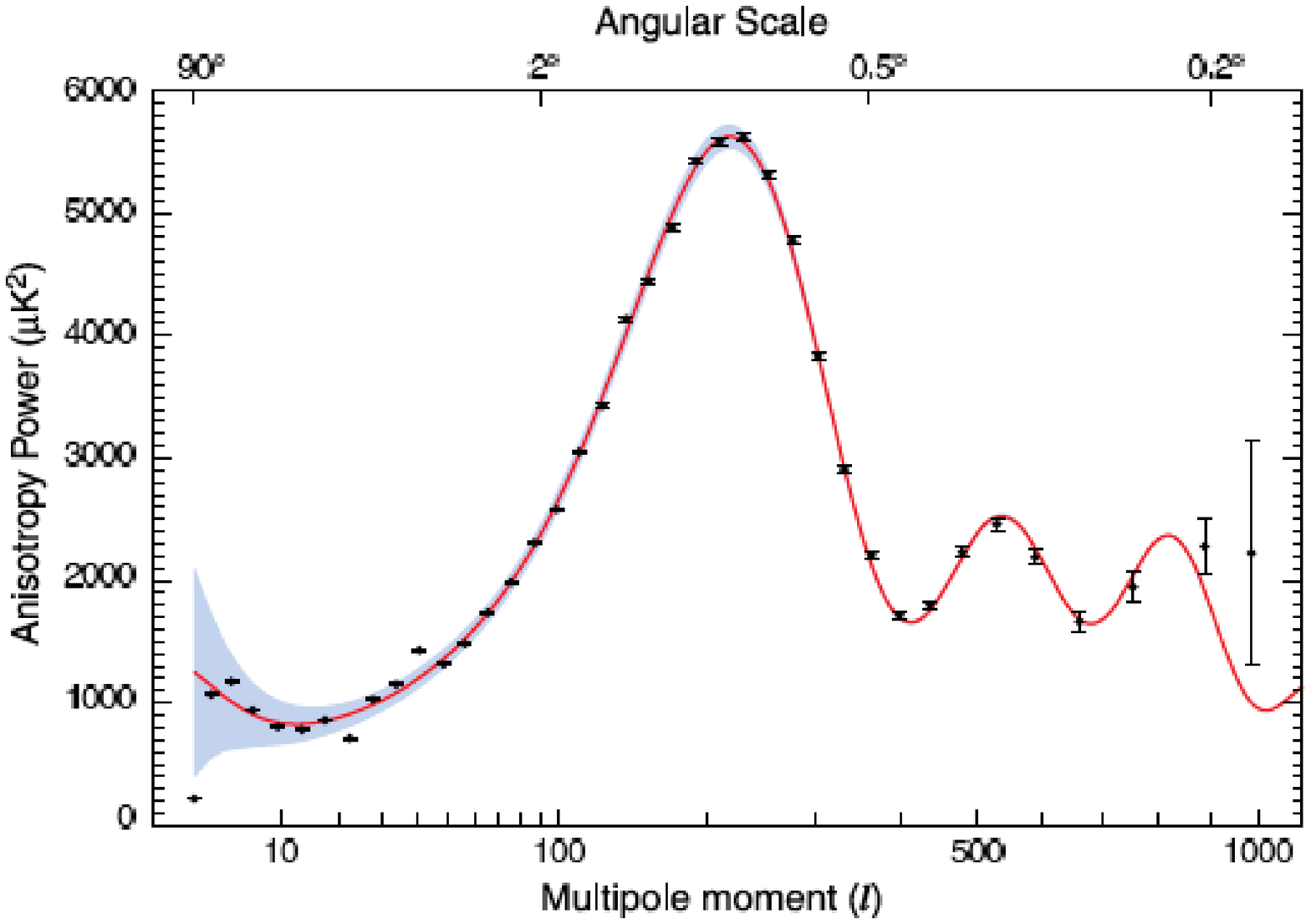, height=1.2in, width=1.2in} & \qquad\quad
\epsfig{file=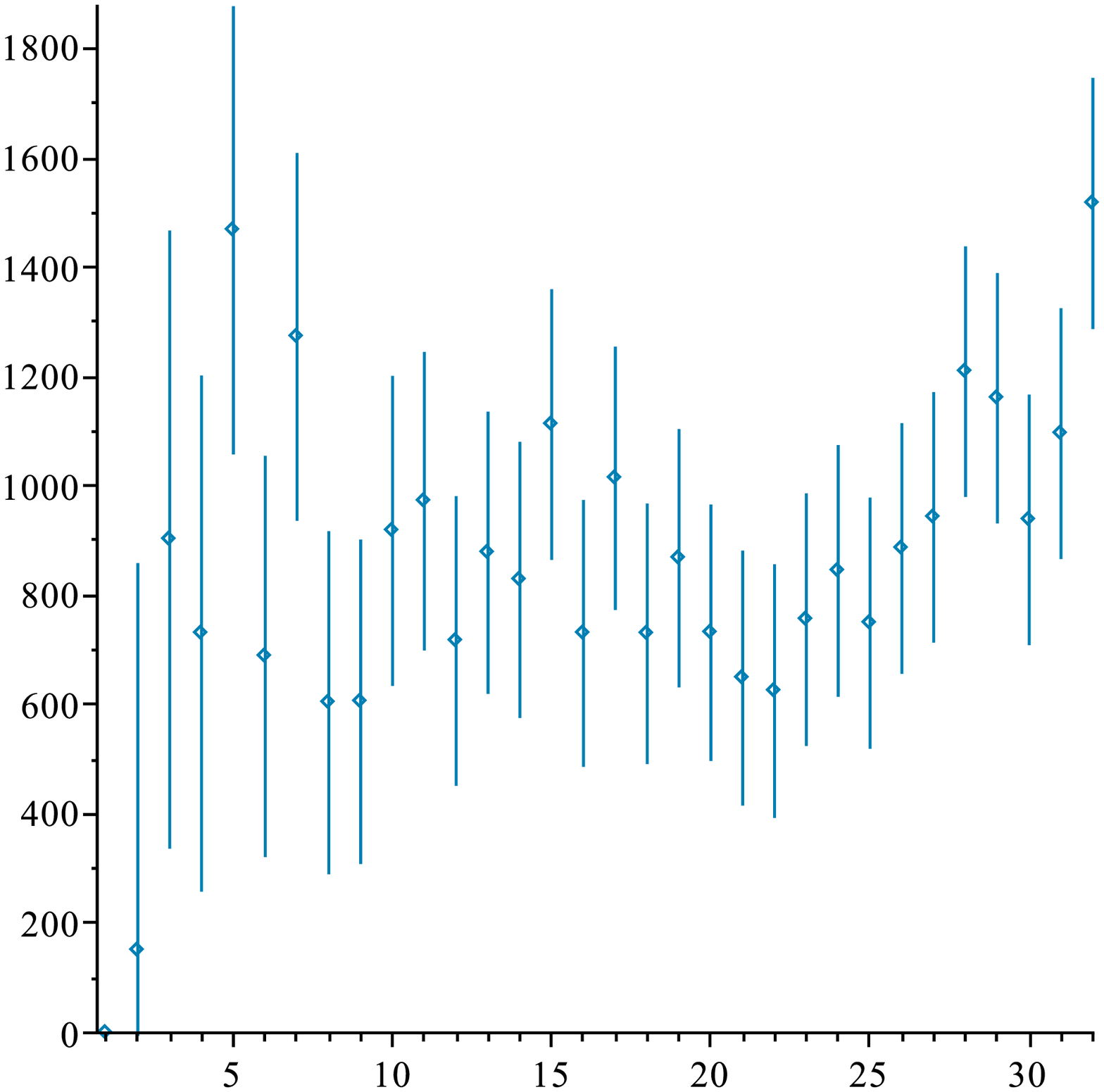, height=1.2in, width=1.2in} & \qquad\quad
\epsfig{file=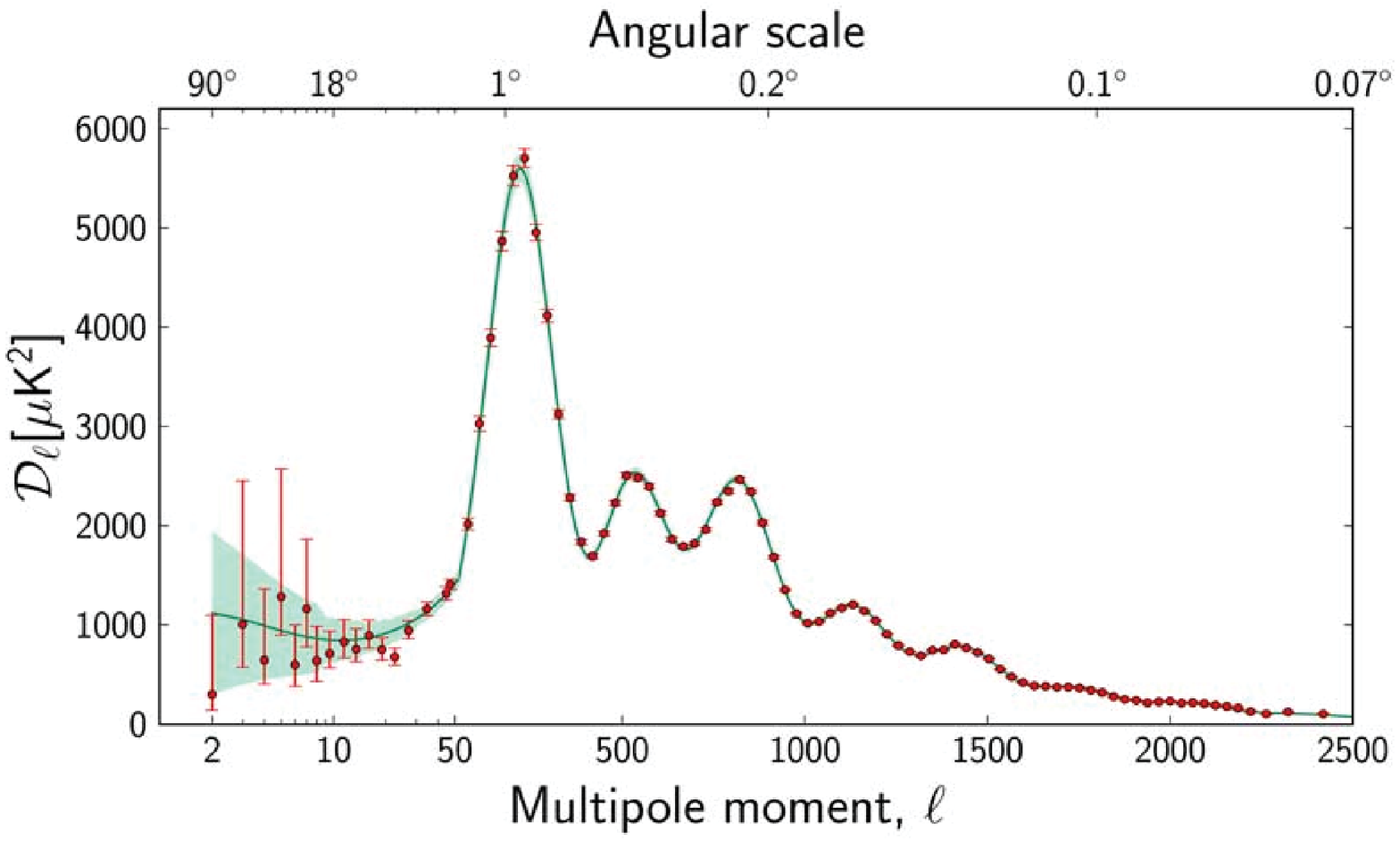, height=1.2in, width=1.2in}
\end{array}$
\end{center}
\caption{\small
WMAP9 determination \cite{WMAP9} of the CMB angular power spectrum (left),
 the raw data for its low--$\ell$ portion (center) and the corresponding PLANCK determination \cite{PLANCK} (right).
The anomalies of interest in this paper concern the low-$\ell$ region detailed in the central portion of the figure, which is based on WMAP9 data,
 while the shadows in the left ends of the outer portions are meant emphasize the role of ``cosmic variance''.
}
\label{fig:WMAP9-PLANCK}
\end{figure}

The recent PLANCK measurements \cite{PLANCK} have strengthened previous WMAP9 results \cite{WMAP9}, providing  additional signs of anomalies at large scales. The low--$\ell$ end of the spectrum shown in fig.~\ref{fig:WMAP9-PLANCK} appears indeed to display a lack of power with respect to the $\Lambda$CDM slow--roll picture, with a sizable reduction of the quadrupole. This would normally spell crisis, were it not for ``cosmic variance''. A cautious attitude is called for, indeed, in attempting to draw conclusions from $2\ell+1$ distinct pieces of information related to different choices of the ``magnetic quantum number'', which are very few for low values of $\ell$. Keeping in mind this important proviso, let us explore the possible lessons of the spectral distortions of Section \ref{sec:powerspectrum}.

The relation between primordial power spectrum and CMB angular power spectrum is particularly neat for $\ell \lesssim 35$ \cite{mukhanov_slow}, where
\be
A_\ell\;(\varphi_0,{\cal M},\delta) \ = \ {\cal M} \ \ell(\ell+1)\ \int_0^\infty \frac{dk}{k} \ {\cal P}_\zeta \big( k , \varphi_0 \big) \, {j_\ell}^2 \big( k \, 10^\delta \big) \, , \label{bessel}
\ee
Here we are emphasizing the dependence on $\varphi_0$ and $j_\ell$ denotes a spherical Bessel function, and nicely enough late time effects are negligible for $\ell \lesssim 35$. The cosmological evolution thus acts like a filter that deforms slightly the primordial power spectrum by smearing it against peaked combinations of Bessel functions.

The first parameter in eq.~\eqref{bessel} is an overall normalization ${\cal M}$, which accounts for various constants entering the relation between the angular power spectrum and the primordial power spectrum of scalar perturbations and for the conversion to the proper units, $\mu K^2$, but ultimately reflects the scale of inflation.
The exponent $\delta$ controls the horizontal displacement of the features present in the power spectra of Section \ref{sec:powerspectrum}
with respect to the main peaks of the Bessel functions. In more physical terms, $\delta$ is a dial that allows a finer tuning between the largest wavelengths that are entering the cosmic horizon at the present epoch and those that exited at the onset of the inflationary phase. Our comparisons with the observed CMB data then rests on a familiar tool,
\be
\chi^2 \ = \ \ \sum_{\ell=2}^{32} \frac{\left(A_\ell\;(\varphi_0,{\cal M},\delta)\ - \ A_\ell^{\rm WMAP9}\right)^2}{\left(\Delta A_\ell^{\rm WMAP9}\right)^2} \ , \label{chi_squared}
\ee
where $A_\ell^{\rm WMAP9}$ are WMAP9 central values and $\Delta A_\ell^{\rm WMAP9}$ are the corresponding errors. Starting from the two--exponential system of eq.~\eqref{potwoexp}, we proceeded as follows:
\begin{itemize}
\item[1. ] for the two--exponential model we explored about 25 choices for  $\varphi_0$, including the sequence $0,-0.25,-0.5,,\ldots,-3.75,-4$,
 in order to make the role of the pre--inflationary peak more transparent. As we have already stressed, our initial choice for $\gamma$ was motivated by the naive correspondence between the mild exponential and the spectral index,
\be
n_s \ - \ 1 \ = \ 3 \ - \ 2\, \nu \ = \ - \ \frac{6\, \gamma^2}{1\,-\,3\,\gamma^2} \ .
\ee
This result would hold exactly for power--law inflation and identifies the value $\gamma \simeq \frac{1}{12} \sim 0.08$ for $n_s \simeq 0.96$ that we have mentioned in the preceding sections;
\item[2. ]
we repeated the analysis for two--exponential potentials with $\gamma$ equal to $0.04$ and $0.02$ in order to explore systems where the observed spectral index is attained at later epochs. Or, if you will, to take a first, indirect, look at exponential ``hard walls'' accompanying \emph{concave} potentials. Lower values of $\gamma$ led to slight improvements, and a less systematic exploration of the Starobinsky--like potentials of eq.~\eqref{starobinsky} gave similar indications.

\item[3. ] We then examined a further modification, slightly negative values of $\gamma$ capable of enhancing slightly the pre--inflationary peaks. This brought about sizable improvements in the fits, but more striking results were recently obtained \cite{ks3} adding to the potential of eq.~\eqref{potwoexp} a small gaussian bump as in eq.~\eqref{potwoexp_gauss} close to the exponential barrier.
\end{itemize}

\begin{figure}[h]
\begin{center}$
\begin{array}{cc}
\epsfig{file=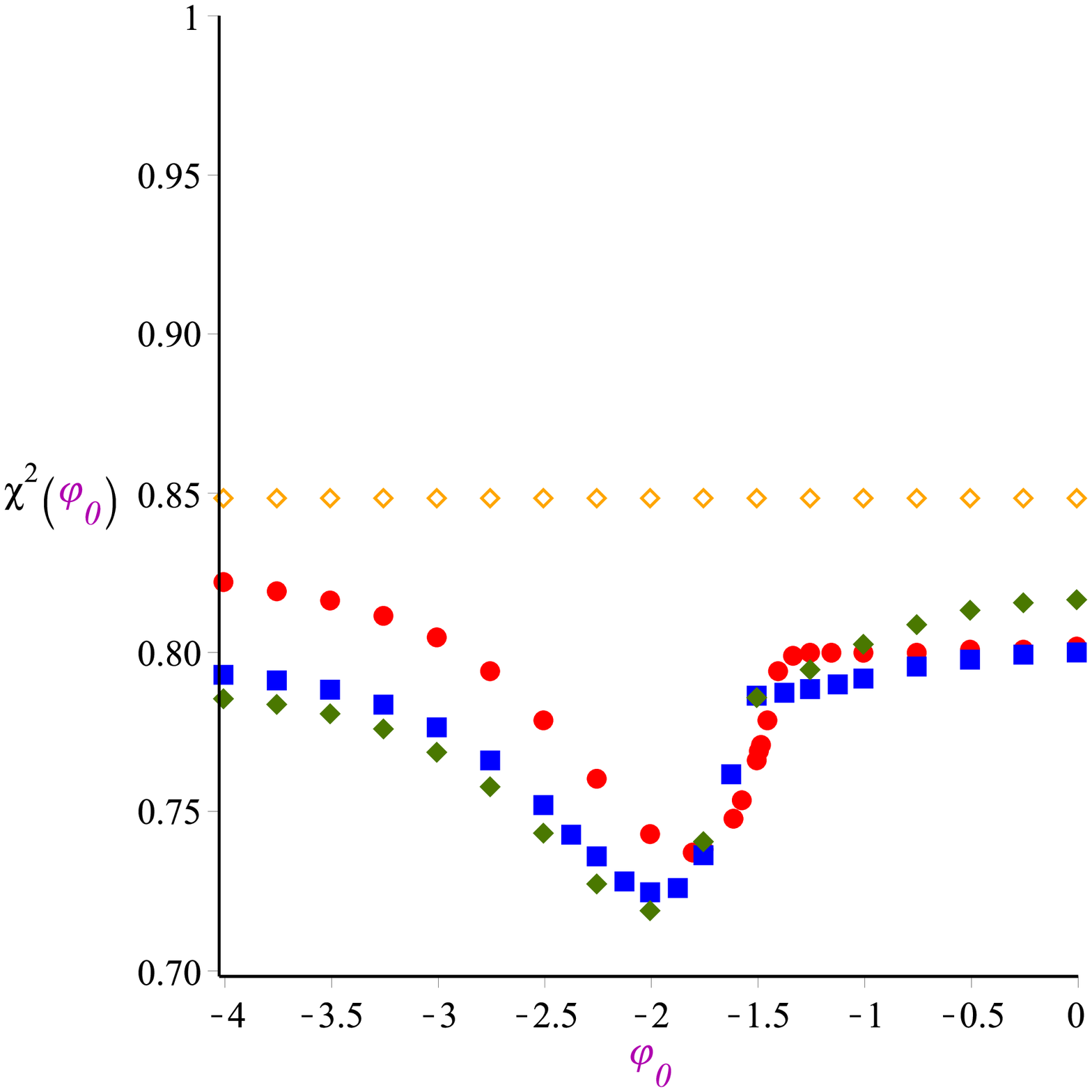, height=1.4in, width=1.4in}& \qquad\qquad
\epsfig{file=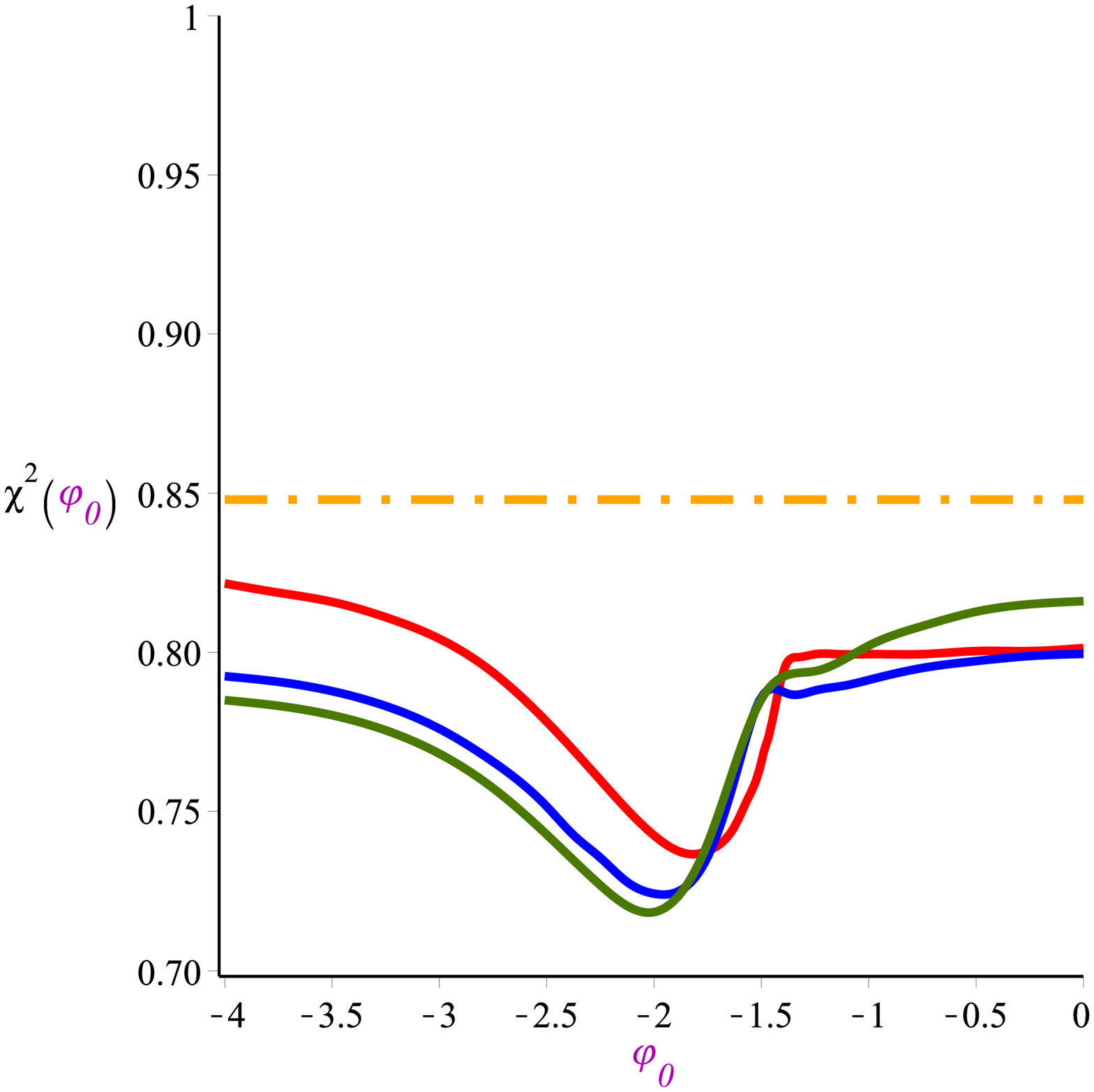, height=1.4in, width=1.4in}
\end{array}$
\end{center}
\caption{\small
Comparisons between WMAP9 raw data and angular power spectra
 for the two--exponential potential of eq.~\eqref{potwoexp},
 in point form (left) and in spline form (right), for $\gamma=0.08$ (red), $\gamma=0.04$ (blue) and $\gamma=0.02$ (green),
 and by the attractor curve (orange).
The minima are $\chi^2_{\rm min}/{DOF}=0.737$, $0.724$ and $0.718$.}
\label{fig:angular_fit_double}
\end{figure}

\begin{figure}[h]
\begin{center}$
\begin{array}{cccc}
\epsfig{file=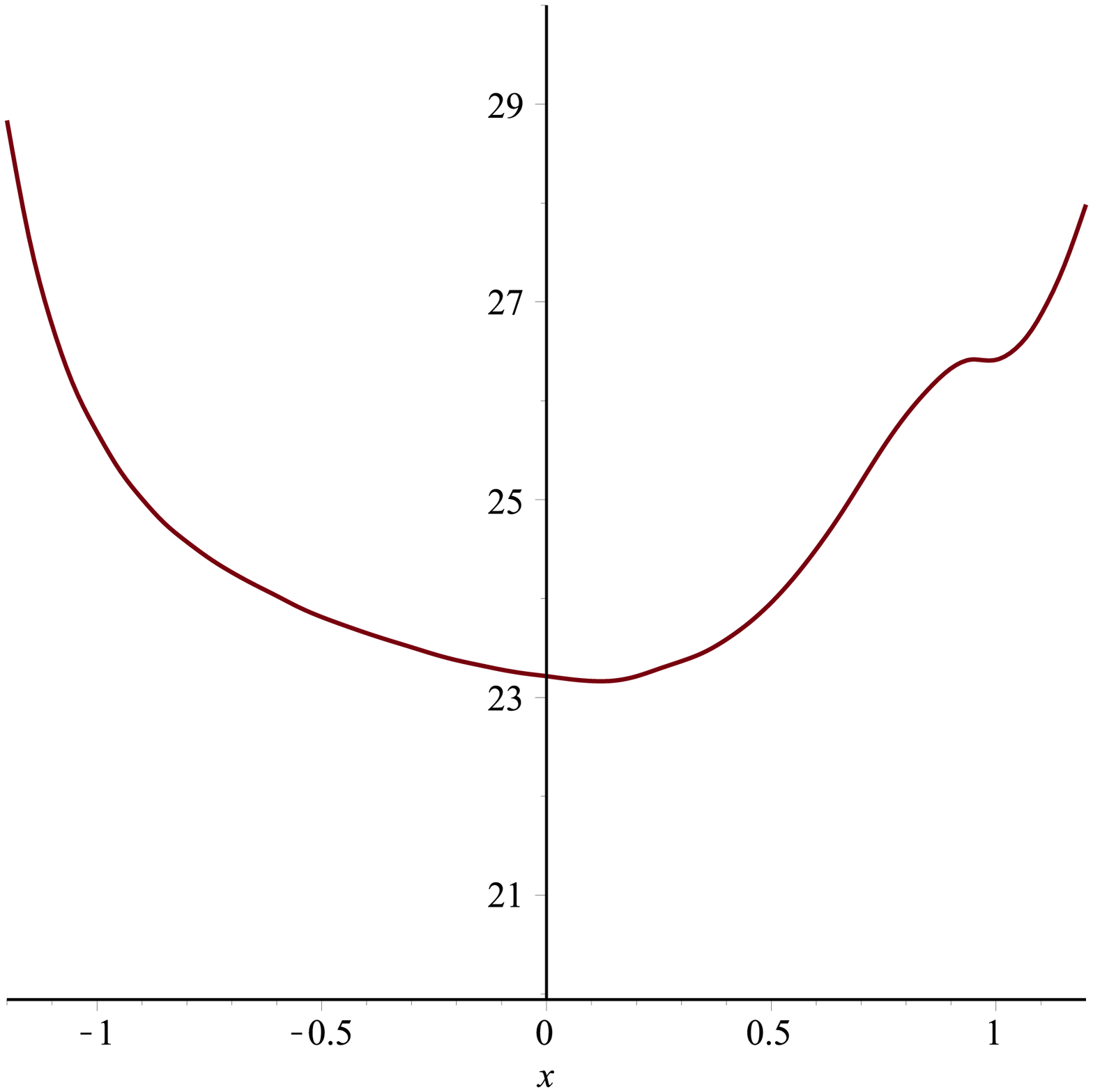, height=1in, width=1in} & \qquad
\epsfig{file=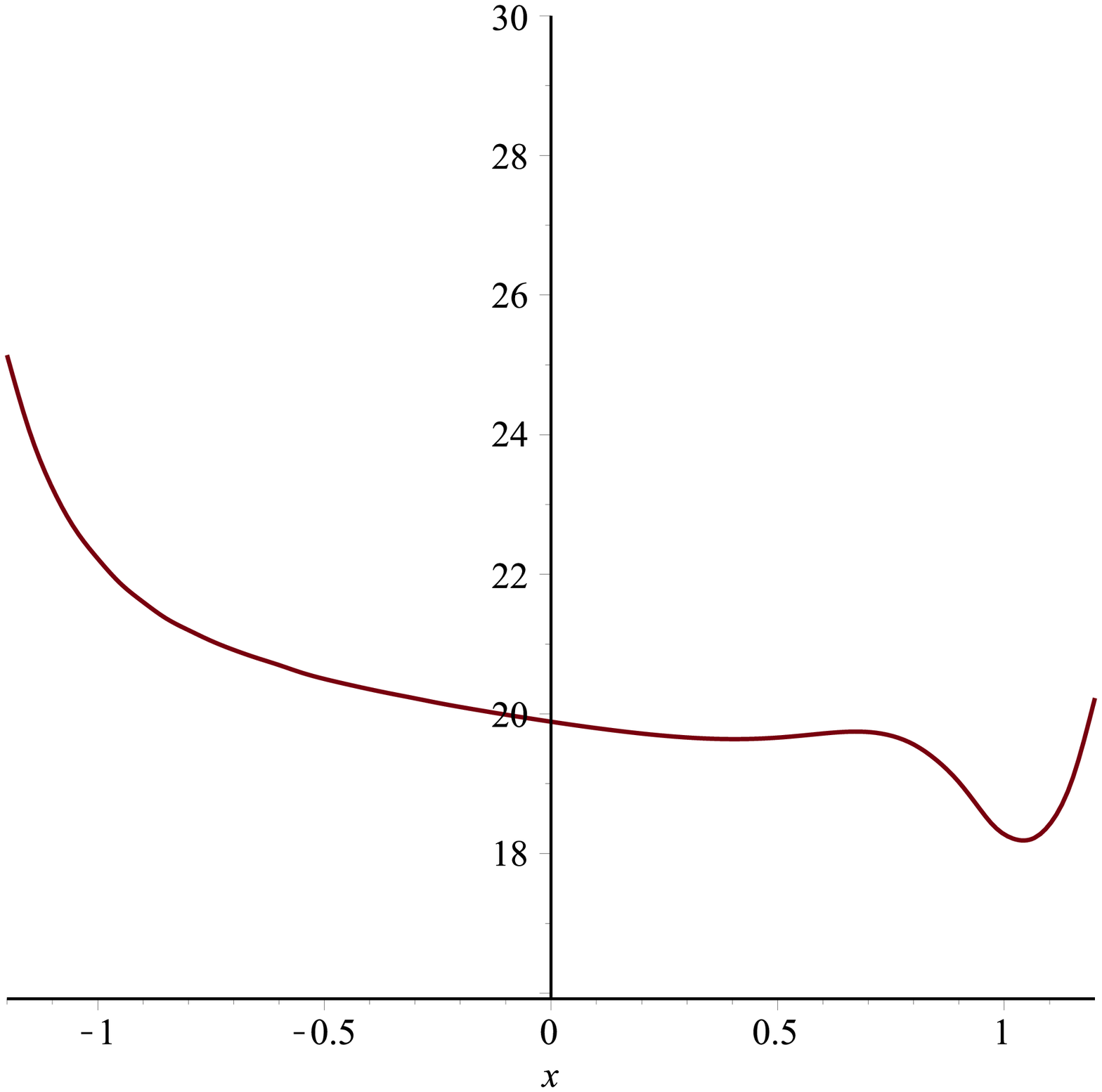, height=1in, width=1in} & \qquad
\epsfig{file=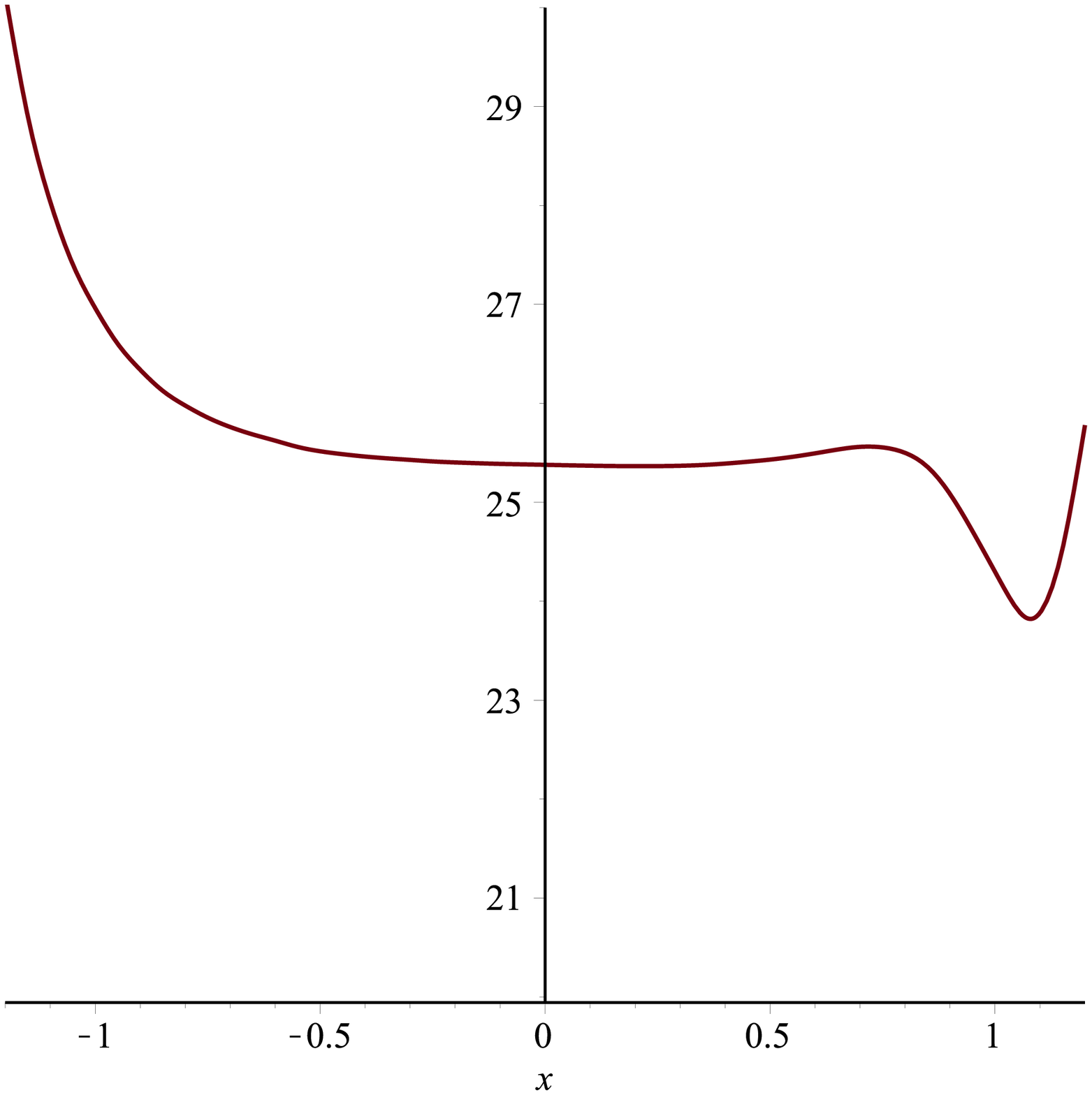, height=1in, width=1in} & \qquad
\epsfig{file=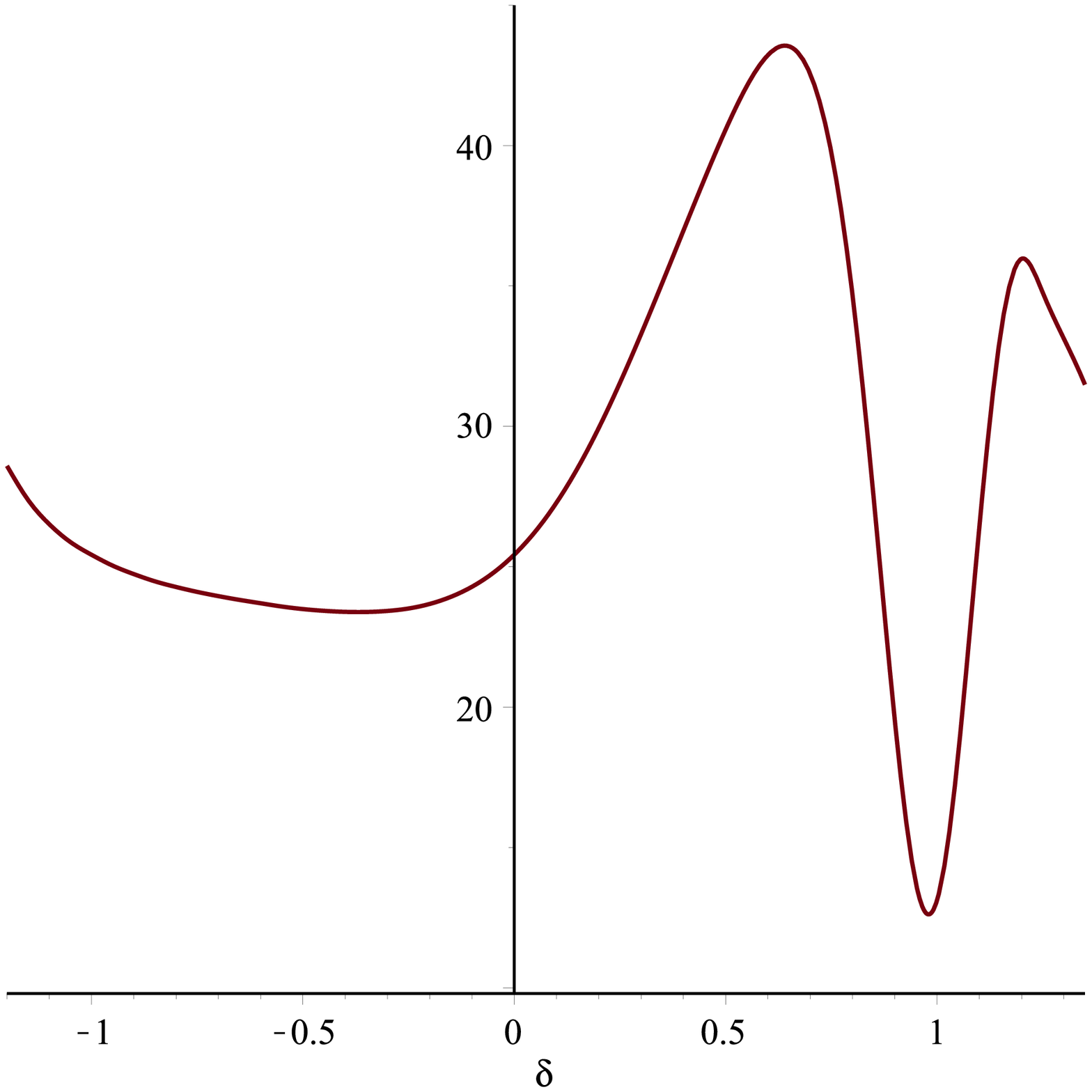, height=1in, width=1in}
\end{array}$
\end{center}
\caption{\small
$\chi^2(\delta)$ for three choices of $\varphi_0$ in the two--exponential model with $\gamma=0.08$ and for the optimal model with a gaussian bump of fig.~\ref{fig:attractor-optimalCl-gaussian}. For $\varphi_0=-1.15$ the peak does not have a sizable effect on the fit, which is dominated by the low quadrupole. The shallow minimum for negative $\delta$ adapts to it the low--frequency depression of the power spectrum (first figure). For $\varphi_0=-1.8$ the peak is well formed and drives the $\chi^2$ to correlate with it features present around $\ell=5$ (second figure). For $\varphi_0=-4$ the peak continues to dominate, but now lies next to the attractor spectrum. This raises the plateau and with it the minimum, since the ${\cal A}_\ell$ display a tendency to grow, rather than to decrease, after the peak (third figure). Finally, for the sake of comparison, the second type of figure for the optimal gaussian bump of fig.~\ref{fig:attractor-optimalCl-gaussian}, with its deeper minimum (last figure).}
\label{fig:angular_fit_double_delta}
\end{figure}

For every choice of $\varphi_0$ two parameters were determined in order to optimize the comparison with WMAP9 raw data:
\begin{itemize}
\item we minimized eq.~\eqref{chi_squared} analytically with respect to the normalization factor ${\cal M}$ present in eq.~\eqref{bessel};
\item we then identified by wide scans optimal choices for the parameter $\delta$ in eq.~\eqref{bessel}.
\end{itemize}
The results of the fits with $\gamma=0.08$, $0.04$ and $0.02$ are collected in fig.~\ref{fig:angular_fit_double}.
These plots display clearly the presence of a transition between regions dominated by the pre--inflationary peak and by the infrared depression. Thus, for instance, the $\gamma=0.08$ plots show initially a mild \emph{decrease} of $\chi^2$ down to a minimum corresponding to  $\chi^2_{min}/DOF \simeq 0.737$, followed by an almost sudden \emph{increase}, and then essentially by a plateau that extends up to $\varphi_0=0$.
This interesting behavior accompanies
 the transition from a pre--inflationary peak lying next to the attractor spectrum, the case well described in \cite{destri} and corresponding to the right portion of fig.~\ref{fig:double},
 to the intermediate pre--inflationary peaks of Section \ref{sec:powerspectrum} and \cite{ks}, and finally to a region where the pre--inflationary peak disappears altogether, leaving only the wide infrared depression discussed in detail in \cite{dkps} (see fig.~\ref{fig:angular_fit_double_delta}).
\begin{figure}[h]
\begin{center}$
\begin{array}{ccc}
\epsfig{file=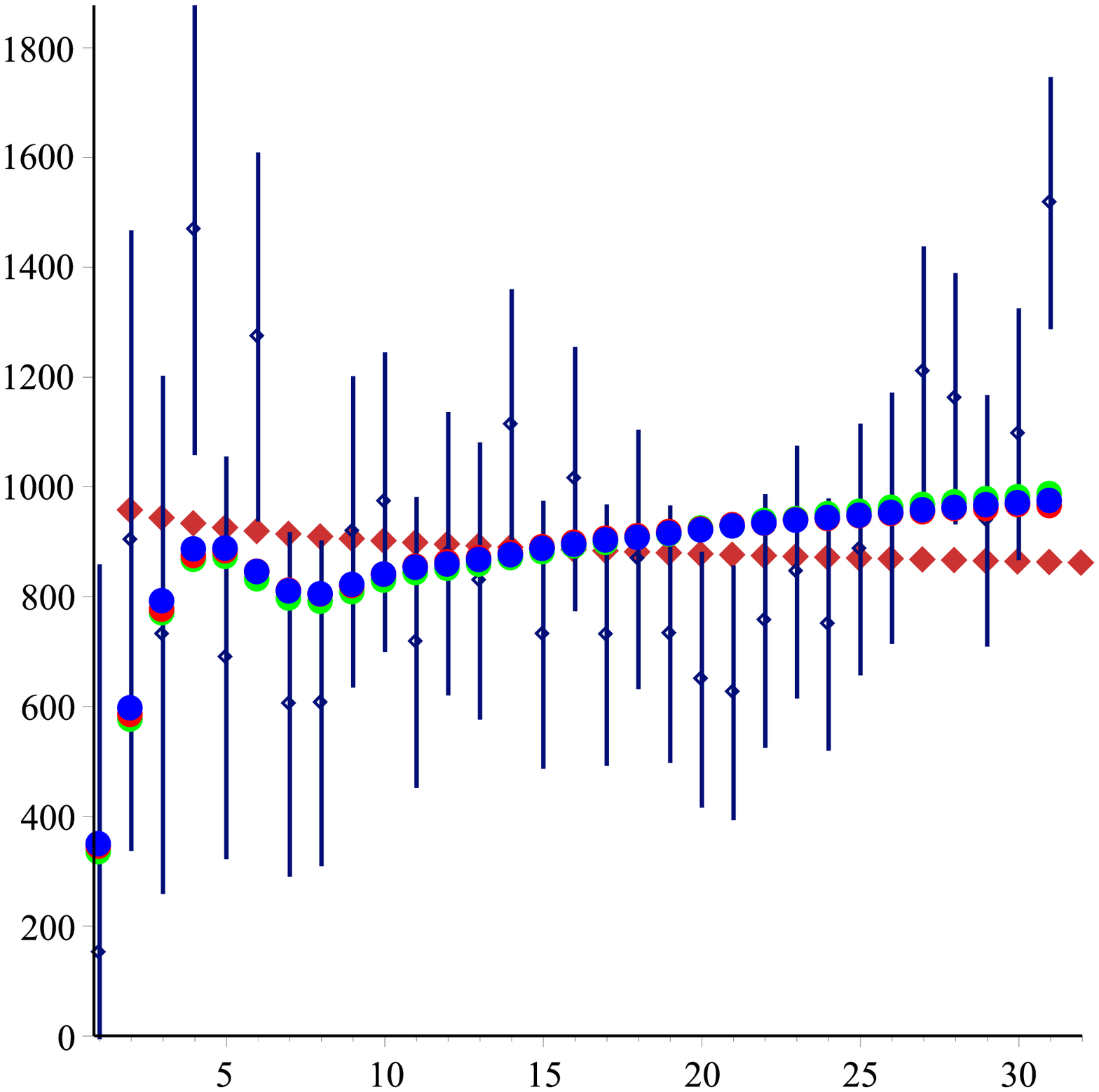, height=1.1in, width=1.1in} & \qquad\quad
\epsfig{file=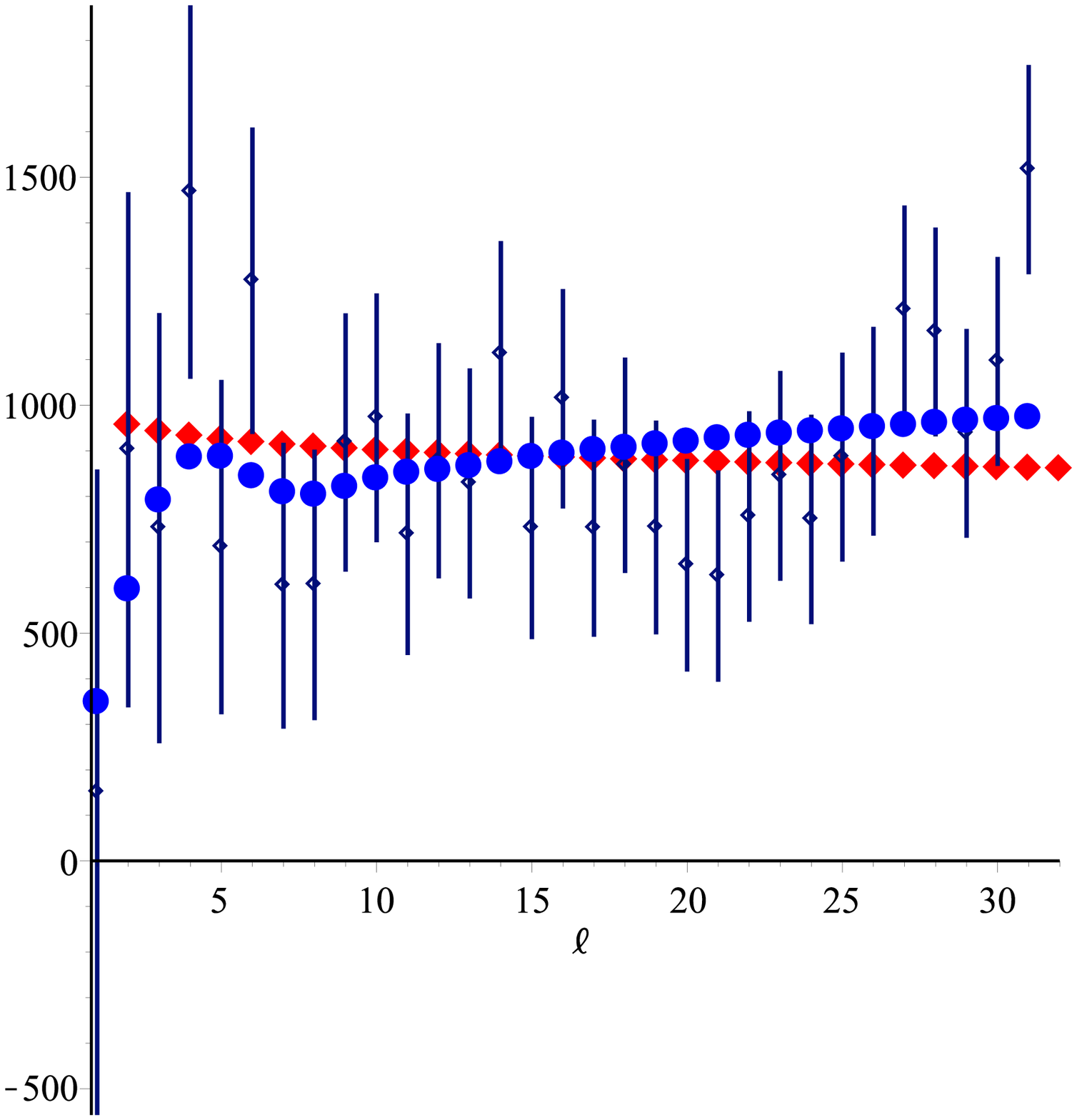, height=1.1in, width=1.1in} & \qquad\quad
\epsfig{file=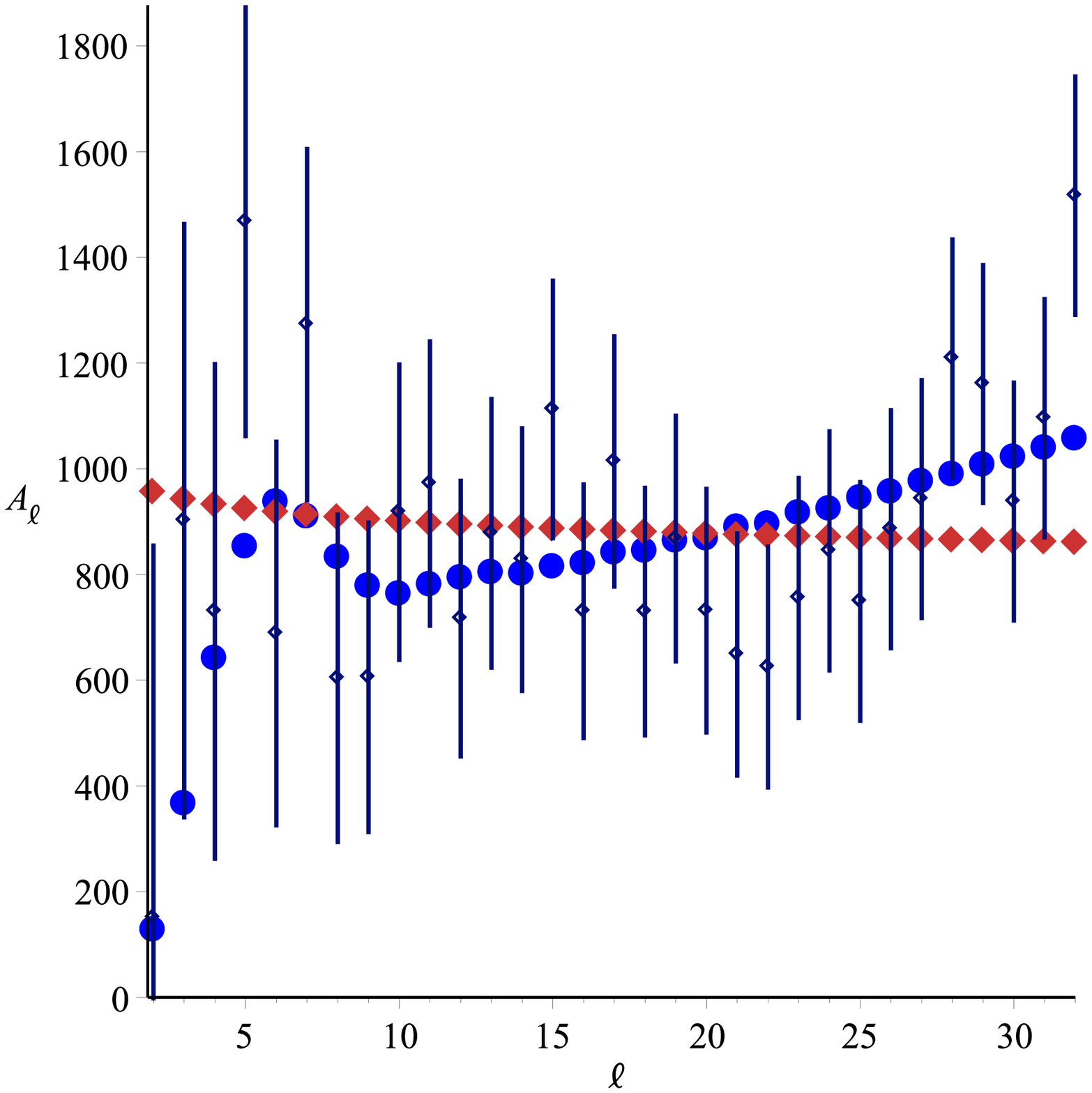, height=1.1in, width=1.1in}
\end{array}$
\end{center}
\caption{\small Optimal $A_\ell$ for $\gamma=0.08$ (blue), the attractor $A_\ell$ (orange) and the WMAP9 raw data (left), and similar plots for $\gamma=0.04$ (center) and for $\gamma=-0.125$ (right). The corresponding values of $\chi^2$ are 21.3, 20.8 and 17.5, to be compared with the attractor value, 25.5. }
\label{fig:attractor-optimalCl}
\end{figure}
\begin{figure}[h]
\begin{center}$
\begin{array}{ccc}
\epsfig{file=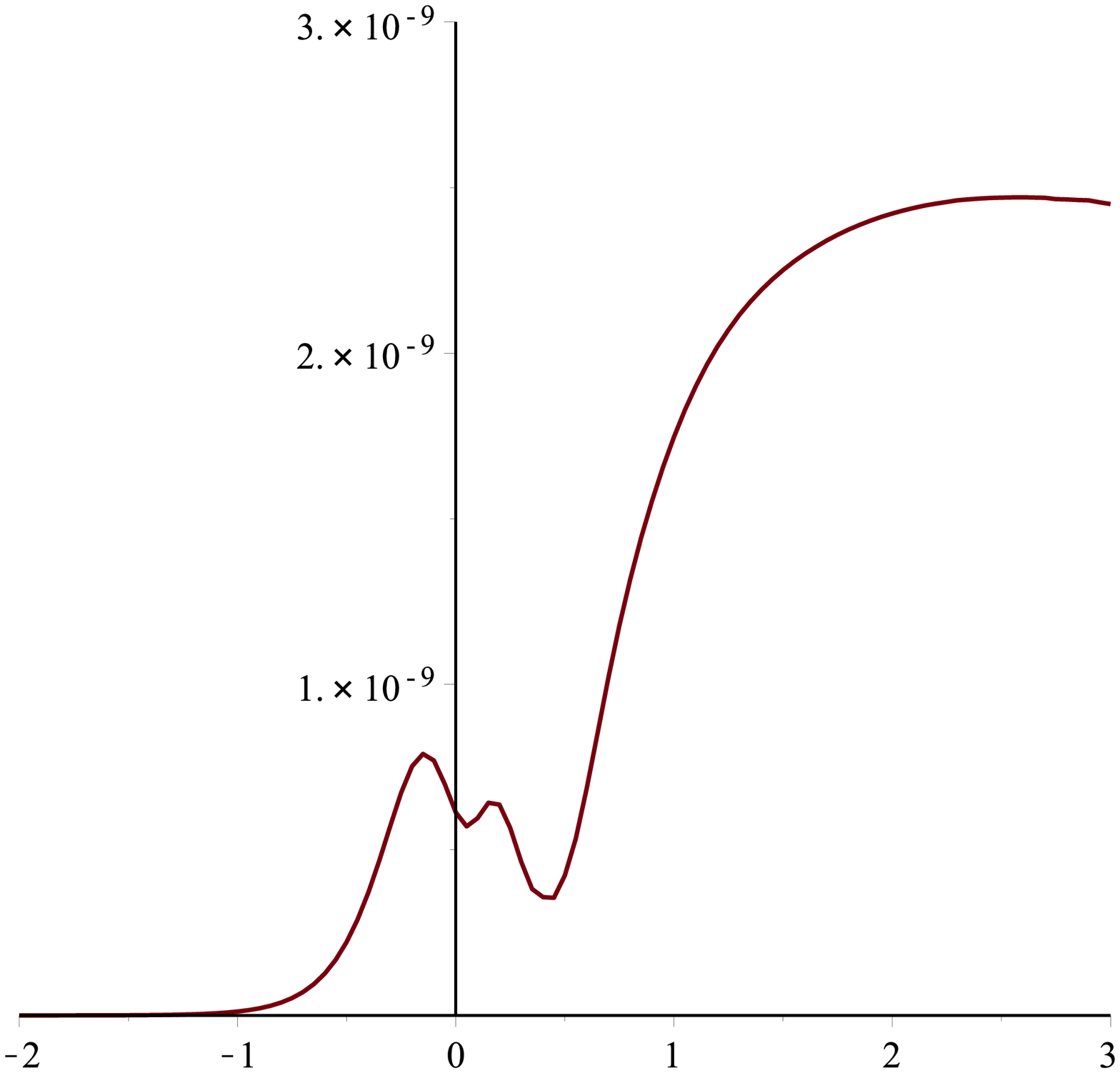, height=1.1in, width=1.1in}& \qquad\quad
\epsfig{file=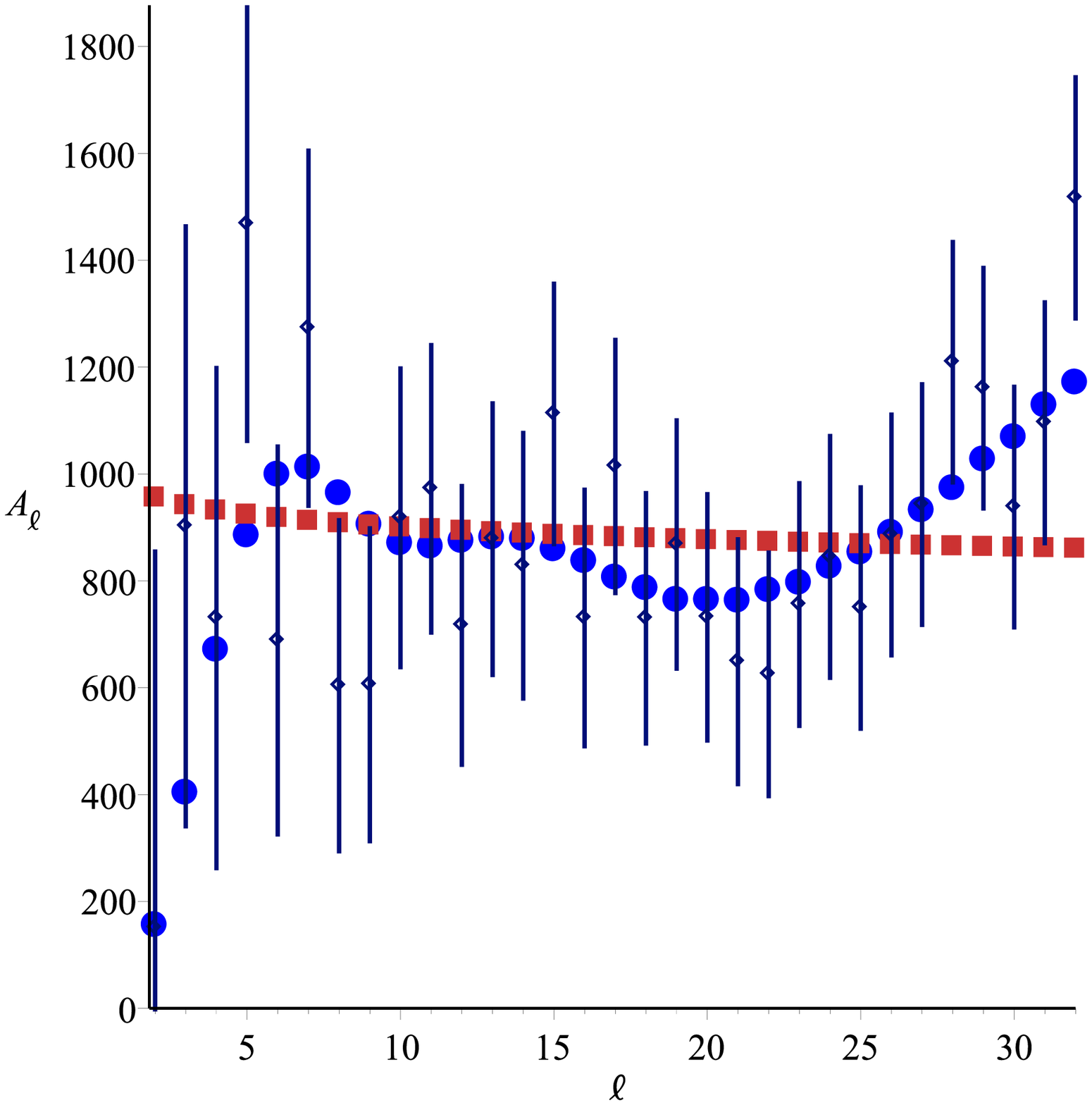, height=1.1in, width=1.1in}& \qquad\quad
\epsfig{file=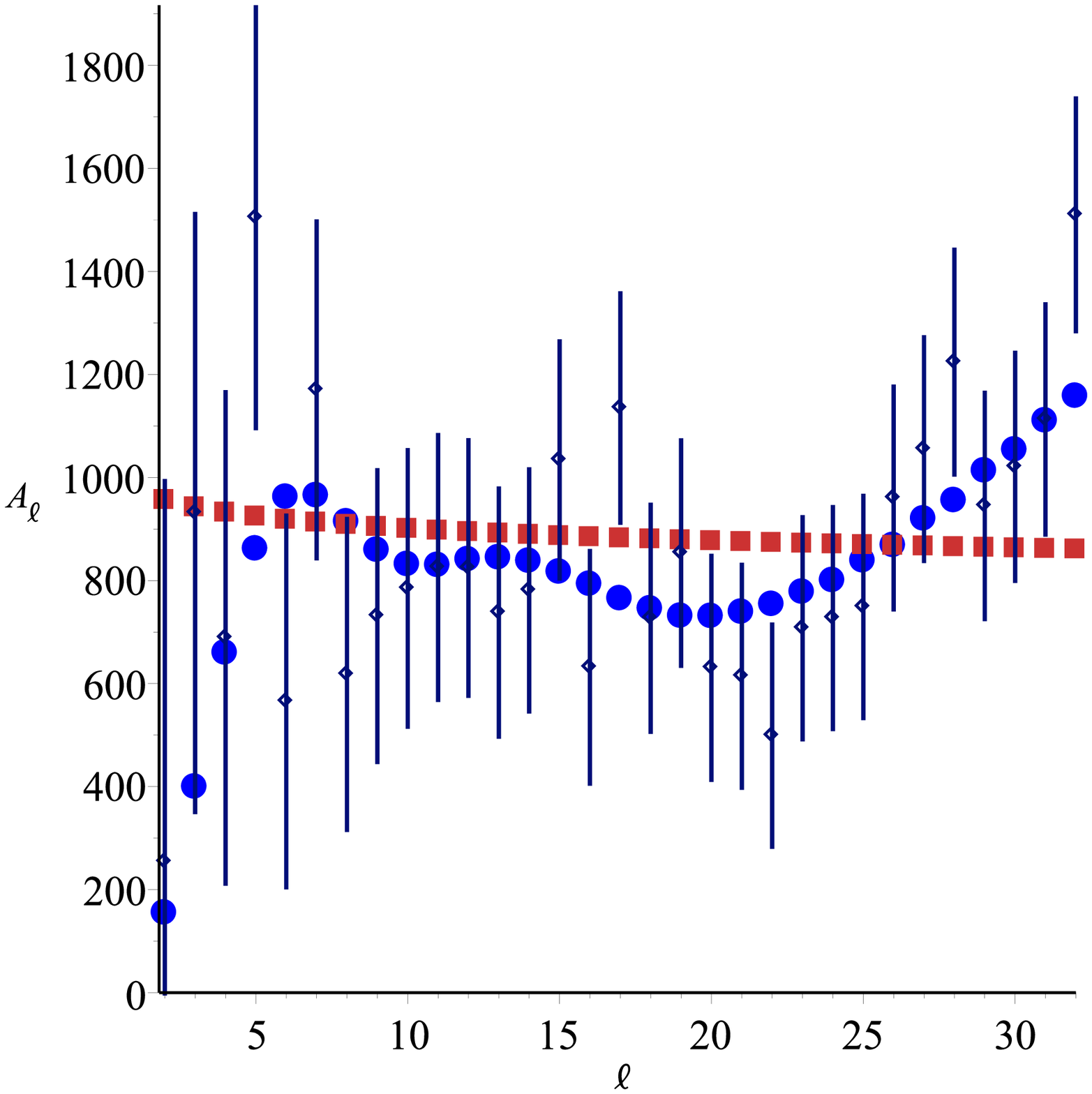, height=1.1in, width=1.1in}
\end{array}$
\end{center}
\caption{\small The best choice that we could find for a power spectrum resulting from a two--exponential system with $\gamma=0.08$ (left) and a small gaussian bump, as in eq.~\eqref{potwoexp_gauss}. The other parameters are $(a_1,a_2,a_3)=(0.065,4,1)$, and $\varphi_0=-0.38$. The low--$\ell$ CMB fit (with blue points for the corresponding $A_\ell$ and red points for the attractor $A_\ell$) with WMAP9 data (center) gave $\chi^2=12.6$, to be compared with an attractor value of 25.5. A similar fit with PLANCK data \cite{PLANCK2} (right) gave $\chi^2=15.6$, to be compared with an attractor value of 30.3.}
\label{fig:attractor-optimalCl-gaussian}
\end{figure}
In fig.~\ref{fig:attractor-optimalCl} we display the optimal two--exponential angular power spectra obtained for $\gamma=0.08$ (the value corresponding to a spectral index $n_s \simeq 0.96$), and for the lower values $\gamma=0.04$ and $\gamma=-0.125$, together with corresponding attractor ${\cal A}_\ell$ and raw WMAP9 data. For the sake of comparison, in fig.~\ref{fig:attractor-optimalCl-gaussian} we also display the best example that we could construct adding a small gaussian bump next to the exponential wall of BSB. \emph{This gives rise to a ``roller coaster'' effect, with the scalar slowing down again, after the bounce, as it overcomes the bump. This slight modification of the two--exponential potential can also add, in the power spectrum, a second smaller peak and, right after the two, a fast growth with an initially convex profile} reminiscent of those encountered for negative $\gamma$, some of which are displayed in fig.~\ref{fig:negative_gamma}. \emph{The corresponding values of $\chi^2$ can be strikingly lower, and we reached a limit of about 12.6, which is about 50\% of the attractor result} \cite{ks3}. Even if we are introducing the three additional parameters $a_1$, $a_2$ and $a_3$ characterizing the bump, the $\chi^2/DOF$  goes down to about 0.48, to be compared with the values around 0.7 that obtain for the two--exponential model. One is clearly fiddling with parameters of phenomenological origin: the only datum extracted from String Theory is the hard exponential, while the mild one is needed to grant an inflationary phase, and finally the bump enhances the effects of the reflection. Therefore, some care should be exercised in assessing the significance of our results, but the bounce, the only datum that is somehow extracted from String Theory, remains the single key input. Let us add that in \emph{all our angular power spectra with better CMB fits, the small peak arising from the reflection by the hard exponential ends up in the angular power spectra around $\ell=5$}, an intriguing fact that could be worthy of further attention.

In conclusion, none of the different models is statistically excluded but
the data display a clear, if slight, preference for the pre--inflationary peaks inspired by BSB. Negative $\gamma$'s improved matters a bit, but a striking improvement resulted when a very simple deformation, a small gaussian bump, was added next to the hard exponential. In this fashion, we reached a 50\% reduction in $\chi^2$ with respect to the attractor value. We hope that an ongoing full--fledged likelihood analysis will help to clarify the import of this preliminary analysis \cite{gkns}.

\section*{Acknowledgments}

We are grateful to D.~Chialva, A.~Gruppuso, P.~Natoli, F.~Nozzoli and G.~Rolandi for stimulating discussions. This work was supported in part by the ERC Advanced Grant n. 226455 (SUPERFIELDS), by Scuola Normale Superiore, by the Tokyo Metropolitan University, by INFN (I.S. Stefi) and by the JSPS KAKENHI Grant Number 26400253. The authors would like to thank for the kind hospitality CERN, Scuola Normale Superiore, the \'Ecole Polytechnique and the Tokyo Metropolitan University.

\end{document}